\documentclass[dvips,11pt]{article}
\usepackage{varioref,exscale,latexsym,amsmath,amssymb}
\usepackage{graphicx}
\usepackage{slashed}
\def\beq{\begin{equation}}

\def\eeq{\end{equation}}

\def\beqa{\begin{eqnarray}}

\def\eeqa{\end{eqnarray}}

\title{{\bf Equivalence Principle Violations and Couplings of a Light Dilaton\\}}
\author{Thibault Damour$^{a}$ and John F. Donoghue$^{a,b}$, \\ \\
$^a$Institut des Hautes \'{E}tudes Scientifiques \\
Bures sur Yvette, F-91440, France\\
and \\
$^b$Department of Physics\\
University of Massachusetts\\
Amherst, MA  01003, USA
 \\}
\begin{document}
\begin{titlepage}
\maketitle
\begin{abstract}
We consider possible violations of the equivalence principle through the exchange of a light `dilaton-like' scalar field.
Using recent work on the quark-mass dependence of nuclear binding, we find that the dilaton-quark-mass coupling induces significant equivalence-principle-violating
effects varying like the inverse cubic root of the atomic number - $A^{-1/3}$. We provide a general parameterization of the scalar couplings, but argue that two parameters are likely to dominate the equivalence-principle phenomenology. We indicate the implications of this framework for comparing
the sensitivities of current and planned experimental tests of the equivalence principle.
\end{abstract}
\vspace{0.2 in}
\end{titlepage}

\section{Introduction}
At the heart of the theory of General Relativity is Einstein's Equivalence Principle (EP). The weak Equivalence Principle predicts the composition independence of the accelerations of test masses in a gravitational field. This has been probed at a present sensitivity of
\begin{equation}
\frac{\Delta a}{a} \sim 10^{-13}
\end{equation}
in innovative and difficult experiments \cite{Schlamminger:2007ht, lunarlaser}.
Further tests of this principle remain important and relevant for new physics \cite{damour,Turyshev:2009ir}.
We are fortunate that there are several initiatives to push the sensitivity several orders of magnitude further using new space-based experiments such as MICROSCOPE \cite{microscope}, the Galileo Galilei project \cite{gg} and
STEP \cite{step} as well as
new types of experiments using cold atoms \cite{cold atoms, mueller} and sub-orbital rockets
 \cite{POEM}.

One possible source of EP violation is a very light\footnote{We will generally assume in the following that the scalar
field we consider is essentially massless on the scales that we discuss.} scalar field with a coupling to matter
that is weaker than gravitational strength. We will refer to these generically as `dilatons', although they may have origins other than
string theory or models involving dilation symmetry. As will become clear below, we will phenomenologically define a `dilaton'
as a scalar field $\phi$ whose couplings to matter effectively introduce a $\phi$ dependence in the basic dimensionless constants
of Nature (such as the fine-structure constant etc.).
String theory may have such scalars in the low energy limit ( string dilaton, moduli), and
these can naturally lead to EP violation at a sizeable level \cite{Taylor:1988nw,Damour:1994ya,runaway,Damour:1992we,Kaplan}. Likewise, theories
of quintessence predict a light scalar, as do theories with continuously varying coupling constants as well as some theories of dark matter. While scalars
lead to an attractive interaction, like usual gravity, they do not couple {\em universally} to  all forms of energy in the same way as in
general relativity. Thus we expect differences in the forces for different elements.

Additionally, independently of any specific theoretical model one might argue (along the `anthropic' approach to the issue of a possibly extremely vast `multiverse' of cosmological and/or string backgrounds) that: (i) the `Equivalence Principle' is not a fundamental symmetry principle of Nature (e.g. it is `violated' in any theory containing very light scalars); (ii) the level $\eta \sim \Delta a / a$ of EP violation can be expected to vary, quasi randomly, within some range {\em of order unity}, over the full multiverse of possible (cosmological and/or theoretical) backgrounds; (iii) as there is probably a maximal level of EP violation, say $0 < \eta_* \ll 1$, which is compatible with the development of life (and of physicists worrying about the EP), one should a priori expect to observe, in our local environment, an EP violation $\eta$ of order of $\eta_*$. It is a challenge to give a precise estimate (or at least upper bound) of $\eta_*$.
We note, however, that this is a scientifically rather well-posed challenge. For instance, one of the necessary conditions for the existence of life is the existence of solar-like planetary systems stable over billions of years. A sufficiently large $\eta \ne 0$ will jeopardize this stability, notably under the influence of external, passing stars. The current very small level of EP violation ensures that stars passing at a distance $D$ disturb the inner dynamics of the solar system only through tidal effects that decrease like $D^{-3}$. An EP violation $\eta$ would increase this disturbing effect to a level
$\propto \eta R^{-2}$. It is also a well-posed question to determine the level $\eta$ which would destabilize the solar system through internal EP-violating gravitational effects.

Independently of these various motivations, our work here will discuss the general type of composition-dependence of EP violation that is entailed by the existence of a light dilaton-like field. The theoretical challenge is to connect the basic couplings of the dilaton Lagrangian to the properties of real atomic systems.

Our work starts in Section 2 with a review of EP violations, and a general parameterization of possible dilaton couplings, Eq. (\ref{Lint}).
Section 3 connects dilaton coupling parameters with the other couplings of the Standard Model, which is preparation for
understanding the effects of the dilaton couplings. Section 4 is our analysis of the effects in nuclear binding, while Section 5 is a summary of the effects within individual nucleons, and Section 6 describes electromagnetic effects.
In Section 7, we collect the results of the previous sections and give a complete treatment of the phenomenology of equivalence principle violations, including comparisons with existing experiments. Section 8 provides a guide to experimental sensitivities for existing and future experiments. Experimenters who are willing to forgo the theoretical development of Section 3-6 can go directly to Section 7-8 or can consult our shorter paper \cite{equivalenceletter} in which we have collected our most phenomenologically useful results. In particular, Section 7.3 contains what is probably the most useful parameterization of our results and Section 7.4 discusses the present experimental constraints. Section 9 is a brief summary.

\section{Formalism}

\subsection{EP violation}

Let us start by recalling that a massless dilaton $\phi$ modifies the Newtonian interaction between a mass $A$ and a mass $B$, into the form
(see, e.g. \cite{Damour:1992we})
\begin{equation}
V=-G\frac{m_A m_B}{r_{AB}}(1+\alpha_A \alpha_B).
\end{equation}
If the dilaton mass is important the second term includes an extra exponential factor $\exp(-m_\phi r_{AB})$.
In this interaction potential, the scalar coupling to matter is measured by the dimensionless factor
\begin{equation}
\label{alphaphi}
\alpha_A = \frac{1}{\kappa^2 m_A}\frac{\partial [\kappa m_A(\phi)] }{\partial \phi} .
\end{equation}
Here,  $\kappa \equiv \sqrt{ 4\pi G}$ is the inverse of the Planck mass\footnote{We use units such that $c=1=\hbar$.} so that the product
$\kappa m_A$ is dimensionless. This ensures that this definition of $\alpha_A $ is valid in any choice of units, even if these units
are such that $\kappa$ depends on  $\phi$ (as in the so-called `string frame'). In the following, we shall generally assume that we
work in the `Einstein frame' where the (bare) Newton constant $G$ is independent of $\phi$.
The above expression for the dimensionless scalar coupling $\alpha_A $ has been written in terms of a canonically normalized
scalar field, with kinetic term [using the signature $(+,-,-,-)$]
\begin{equation}
{\cal L}_\phi = \frac12 (\partial \phi)^2 + \cdots
\end{equation}
 Evidently, a small mass term for the dilaton can readily be added if desired.
It can also be convenient to work with the {\em dimensionless} scalar field
\begin{equation}
\varphi \equiv \kappa \phi,
\end{equation}
whose kinetic term is related to the Einstein-Hilbert action via
\begin{equation}
 - \frac{1}{16 \pi G} (R - 2 (\partial \varphi)^2)
\end{equation}
When using $\varphi$ the definition of the dimensionless scalar coupling reads
\begin{equation}
\label{alphavarphi}
 \alpha_A  = \frac{\partial \ln[\kappa m_A(\varphi)] }{\partial \varphi}.
\end{equation}

In terms of the $ \alpha_A$'s, the violation of the (weak) EP , i.e. the fractional difference between the accelerations of two bodies $A$ and $B$
falling in the gravitational field generated by an external body $E$, reads
\begin{equation}
\label{da/a}
\left( \frac{\Delta a}{a} \right)_{AB}\equiv 2\frac{a_A-a_B}{a_A+a_B}=\frac{(\alpha_A - \alpha_B)\alpha_E}{{1+ \frac12(\alpha_A + \alpha_B)\alpha_E}}
\simeq (\alpha_A- \alpha_B)\alpha_E.
\end{equation}
In the last (approximate) equation we have assumed that the $\alpha$'s are small, so that one can neglect the term $\frac12(\alpha_A+ \alpha_B)\alpha_E$
in the denominator.

Our aim here is to provide a general analysis of the possible EP violations in experiments comparing the free fall accelerations of
atoms (and/or nuclei). Most of the effort needed for such an analysis is now understood \cite{Damour:1994ya,runaway, Kaplan,dent}, and we will use it below.
However, one aspect of this analysis has been far less well-studied and understood, namely the contribution to EP violation coming from the
possible $\phi$-dependence of the  {\em nuclear binding energy}. The aim of this paper will mainly be to assess the form of this
contribution, coming from the quark mass contribution to nuclear binding\footnote{Damour \cite{damour}  and Dent  \cite{dent} have highlighted this
need for the study of the nuclear binding energies.}.
Actually, our conclusion will be that this contribution is, possibly in competition with Coulomb-binding effects, likely to dominate the atom-dependence of the EP violation signal (\ref{da/a}).

To motivate our general analysis, let us start by noting that the mass of an atom can be decomposed as
\begin{equation}
\label{massA}
m({\rm Atom}_A) = m_A= m_A^{\rm rest \, mass} + E^{\rm binding}
\end{equation}
where
\begin{equation}
\label{restmassA}
 m_A^{\rm rest \, mass}= Z m_p +N m_n+ Z m_e
\end{equation}
is the rest-mass contribution to the mass of an atom ($Z$ denoting the atomic number and $N$  the number of neutrons),
and where $E^{\rm binding} $ is the binding energy of the atom, which is dominated by the binding energy of the nucleus.
 $E^{\rm binding} \equiv E_3 + E_1$ is the sum of a strong interaction contribution, say $E_3$, and of an electromagnetic one, say $E_1$
(which is dominated by the electromagnetic effect within the nucleus).
The indices $3$ and $1$ are used here as reminders of the gauge groups underlying the considered interactions: namely, $SU(3)$ and $U(1)$.
Note that the index $A$ in $m_A$ is used here (like in the definition of the scalar coupling $\alpha_A$) as a label for distinguishing
several different atoms. It should not be confused with the mass number (or nucleon number) $A \equiv Z+N$ which we shall use below.

\subsection{The general dilaton Lagrangian}

The basic organizing principle that we shall use in our discussion is to keep track of the effect of all the possible $\phi$ modifications
of the terms entering the effective action describing physics at the scale of nuclei in their ground states. We have in mind here
an energy scale $\mu \sim 1$ GeV. At such a scale, one has integrated out not only the effect of weak interactions, but also the heavy
quarks $c, b$ and $t$.  The issue of the possible $\phi$ sensitivity of effects linked to the strange quark $s$ is more delicate. In the Appendix
we argue that the possible EP violations linked to the $\phi$ couplings to $s$ are expected to be quite small. In the bulk of the text we shall
therefore ignore $s$ (assuming that its effect is taken into account by changing some of the quantities we discuss, notably the QCD energy scale
$\Lambda_3$).

In this approximation, we are therefore talking about an effective action containing, as real particles, the electron $e$, the $u$ quark,
and the $d$ quark, with interactions mediated by the electromagnetic ($A_\mu$) and gluonic ($A^A_\mu$) fields. [Here we shall use a rescaled $U(1)$
gauge potential, which incorporates the electron charge $e$, but an unrescaled gluonic field, which does not incorporate the $SU(3)$ gauge coupling $g_3$.]
Then each of the five terms in this effective action, say
\begin{equation}
\label{action}
{\cal L}_{\rm eff} = - \frac{1}{4e^2} F_{\mu\nu}F^{\mu\nu} - \frac{1}{4} F^A_{\mu\nu}F^{A\mu\nu}
+ \sum_{i= e,u,d} \left[ i {\bar \psi}_i {\slashed{D}}(A, g_3 A^A) \psi_i - m_i{\bar \psi}_i\psi_i \right] \; ,
\end{equation}
(where $D(A)$ denotes the Dirac operator coupled to the gauge field(s) $A$)
can couple to $\varphi= \kappa \phi$ with a (dimensionless) coefficient. [We assume that we work in the Einstein frame, with the gravity
and $\phi$ kinetic terms displayed above.] This introduces {\em five} dimensionless dilaton coupling
coefficients, say $d_e, d_g$ for the couplings to the electromagnetic and gluonic field terms, and $d_{m_e}, d_{m_u}, d_{m_d}$
for the couplings to the fermionic mass terms\footnote{We are using here the fact that a $\phi-$dependent coupling to the kinetic term
of a fermion, $f(\phi) {\bar \psi} i {\slashed{D}} \psi$, can be absorbed in a suitable $\phi-$dependent rescaling of $\psi$.}. We shall
normalize these five dimensionless dilaton coupling coefficients $d_e, d_g, d_{m_e}, d_{m_u}, d_{m_d}$ so that they
correspond (when considering the linear couplings to $\phi$) to the following interaction terms
\begin{equation}
\label{Lint}
{\cal L}_{{\rm int} \phi} =  \kappa \phi \left[ + \frac{d_e}{4e^2} F_{\mu\nu}F^{\mu\nu}
-\frac{d_g\beta_3}{2g_3} F^A_{\mu\nu}F^{A\mu\nu} - \sum_{i=e,u,d} (d_{m_i}+\gamma_{m_i}d_g) m_i{\bar \psi}_i\psi_i
\right]
\end{equation}
We shall explain below the notation and our choice of normalization for these interaction terms.

There are two equivalent ways of thinking about the computation of the scalar-matter coupling $\alpha_A$, Eq. (\ref{alphaphi}).
One way is to think that it is given by the matrix element (in the quantum state of an atom) of the operator in the quantum Hamiltonian
(associated to the interaction Lagrangian above)
which is linear in $\phi$. A second way is to think that it is obtained by the chain rule as
\begin{equation}
\alpha_A =  \frac{\partial \ln[\kappa m_A(\varphi)] }{\partial \varphi} =
\sum_a \frac{\partial \ln[\kappa m_A(k_a)] }{\partial k_a}  \frac{\partial k_a }{\partial \varphi} .
\end{equation}
where $\kappa m_A(k_a)$ is the expression of the dimensionless mass ratio $\kappa m_A = m_A/m_{\rm Planck}$ as a
function of the dimensionless coupling constants of Nature, say $k_a= k_1, k_2, \ldots ,k_{20}$, entering the
Standard Model. Actually, because of the limited number of terms entering the relevant low-energy
action (\ref{action}), there are only five relevant dimensionless constants of Nature $k_a$ corresponding to the
five terms in (\ref{action}). As we shall see in detail below, the five terms in the interaction terms (\ref{Lint}) precisely correspond to
introducing a  $\phi$ dependence in the
five following dimensionless constants of Nature,
\begin{equation}
\label{constants}
\alpha, \kappa \Lambda_3, \kappa m_e, \kappa m_u, \kappa m_d,
\end{equation}
where $\alpha= e^2/ (4 \pi)$ is the fine-structure constant, $\Lambda_3$ the QCD energy scale, $m_e$ the electron (pole) mass,
and where $m_u$ and $m_d$ denote some renormalization-group-invariant measures of the light quark masses (say, the $\mu$-running masses
taken at the multiple of $\Lambda_3$ which is equal to $1$ GeV).

In the next Section we shall relate our normalisation of the five dimensionless dilaton coupling parameters $d_a$ entering (\ref{Lint})
to the constants (\ref{constants}), and
explain in more detail the dependence of the mass of an atom on the five constants (\ref{constants}), and thereby on the five
dilaton parameters  $d_a$.

\section{Relation between the dilaton coupling parameters $d_a$ and the `constants of Nature'.}

By comparing the $\phi$-interaction Lagrangian (\ref{Lint}) to the other terms in the effective action (\ref{action}), we see that the
meaning of the dilaton coupling coefficients $d_a= d_e, d_g, d_{m_e}, d_{m_u}, d_{m_d}$ seems
clear for four of them. [Actually, we shall see below that the meaning of the quark-mass
couplings $d_{m_i}$ is more subtle, because of the renormalization group running of the quark masses, which is associated with the
$ \gamma_{m_i}d_g$ term in (\ref{Lint}).]
First, the coupling $d_e$ to the electromagnetic field modifies the Maxwell action according to
\begin{equation}
\label{EM}
{\cal L}_{EM} = -\frac{1 - d_e\kappa \phi}{4e^2} F_{\mu\nu}F^{\mu\nu}
\simeq -\frac{1 }{4 (1 + d_e\kappa \phi)e^2} F_{\mu\nu}F^{\mu\nu}
\end{equation}
where the last equality is valid at the linear level in $\kappa \phi$ (which is the level at which we
define the dilaton couplings here).
As we work with a rescaled  electromagnetic field ($ A^{\rm here} = e A^{\rm usual}$), the only location where the electric charge occurs in the Lagrangian is
the one explicitly shown above. This allows the dilaton field to be absorbed into the following $\phi$ dependence of the fine-structure constant
\begin{equation}
\alpha(\phi) =   (1 + d_e\kappa\phi ) \alpha  = (1 + d_e \varphi)  \alpha.
\end{equation}
Second, comparing (\ref{Lint}) to the mass terms of the electron and the light quarks, we see that our normalization is such that
 $d_{m_e}, d_{m_u}, d_{m_d}$ introduce the following $\phi$ dependence of the $e, u$ and $d$ masses:
\begin{equation}
 m_i(\phi) = (1 + d_{m_i} \kappa \phi) m_i = (1 + d_{m_i} \varphi) m_i,  \, (i=e,u,d).
\end{equation}

 On the other hand, the terms in (\ref{Lint}) that depend on our `dilaton-gluon'
 coupling $d_g$ call for a more subtle explanation. The choice of these coupling terms
 is such that the coefficient of $d_g$ is invariant under the renormalization-group (RG).
 As the coefficient of $d_{m_i}$ (i.e. the mass term $m_i {\bar \psi}_i\psi_i$)
 is also, separately, RG-invariant, our choice of normalization of the coefficients in
 (\ref{Lint}) gives a RG-invariant meaning to both $d_g$ and the $d_{m_i}$'s.\footnote{We are
 here talking about invariance under the QCD-driven running of the QCD gauge coupling $g_3$,
 and of the masses of fermions coupled to QCD. In view of the smallness of the electromagnetic
 coupling $\alpha \simeq 1/137 \ll \alpha_3$, we are neglecting the RG-running driven by electromagnetic effects. If one wanted to take it into account, one should add to
 (\ref{Lint}) additional terms linked to the QED trace anomaly.}

\subsection{Connection with the QCD trace anomaly}

 The phenomenological consequences (for the scalar coupling to hadrons) of the RG-invariant nature of the
 couplings in (\ref{Lint}) can be seen in two (equivalent) ways. One way (which was used
 by \cite{SVZ78} and \cite{Kaplan}) consists in remarking that the definition of the $d_g$-dependent
 terms in (\ref{Lint}) is such that they couple $\phi$ to the anomalous part of the
 trace of the gluon stress-energy tensor, namely
\begin{equation}
\label{Lintg}
 {\cal L}_{g \phi}= - d_g \kappa \phi  T_{g}^{\rm anom}
\end{equation}
where \cite{CollinsDuncanJoglekar77}
\begin{equation}
\label{Tg}
 T_{g}^{\rm anom} = \left[ \frac{\beta_3}{2g_3}  F^A_{\mu\nu}F^{A\mu\nu} + \gamma_m \sum_i m_i {\bar \psi}_i\psi_i \right]_{\mu}
\end{equation}
Here, $\beta_3(g_3)=\mu \partial g_3/\partial \mu$ denotes the $\beta$ function for the
running of the QCD coupling $g_3$ with the (Wilsonian) sliding energy scale $\mu$,
$\gamma_m(g_3)=-\mu \partial \ln m/\partial \mu$ (with a minus sign on the r.h.s.) is the
(universal) anomalous dimension giving the energy-running of the masses of the QCD-coupled fermions,
and the subscript $\mu$ at the end indicates that the operator on the r.h.s. must be
renormalized at the running scale $\mu$. We recall that, classically, the trace of
the gluonic stress-energy tensor vanishes (because of the conformal invariance of the
Yang-Mills action), but that quantum effects linked to the necessity of regularizing the UV infinities in the product of  gluon field strengths at the
same spacetime point $x$ introduce the (finite) `conformal anomaly' (\ref{Tg}) \cite{CollinsDuncanJoglekar77}. Then, by using the quantum version of
the virial theorem\footnote{We recall that this theorem says that the space integral of the spatial components of the total stress-energy
tensor $T_{\rm tot}^{\mu \nu}= T_{g}^{\mu \nu} +  T_{EM}^{\mu \nu} +  T_{\rm matter}^{\mu \nu}$ vanishes in an equilibrium bound state.},
one can see \cite{SVZ78,Kaplan} that the coupling
(\ref{Lintg}) means that $d_g$ measures the coupling of $\phi$ to the part of the total
mass-energy of the considered hadron which is due to the (renormalized) gluonic
field energy, say  $ M_g $ (where $ M_g $ can be defined by subtracting from the
total mass both the non-anomalous mass-term contributions
$\langle \sum_i m_i {\bar \psi}_i\psi_i \rangle$, and the electromagnetic one).

\subsection{Renormalization group analysis}

A second way of discussing the
 consequences (for the scalar coupling to hadrons) of our normalization of
 couplings in (\ref{Lint}) is phenomenologically illuminating. It consists in noting that our
 RG-invariant definitions are equivalent to very simple consequences for the $\phi$
 dependences of both the QCD mass scale $\Lambda_3$, and the values of the quark masses
 at the scale $\mu=\Lambda_3$. [Note that both $\Lambda_3$ and $m_i(\Lambda_3)$ are
 RG-invariantly defined quantities.] Let us start by defining the QCD mass scale $\Lambda_3$
 as being the mass scale at which the running QCD coupling $g_3(\mu)$ reaches some
 fixed, reference dimensionless number of order unity, say $g_*=2.5$. [This numerical
 value, which corresponds to $\alpha_*=g_*^2/(4 \pi) =0.5$, is approximately reached when
 the running scale $\mu \simeq 1$ GeV (see, e.g., the figure giving $\alpha_s(\mu)$ in
 the QCD review in \cite{RPP}).] This definition of $\Lambda_3$ can be re-expressed in terms of the value
 $g_c \equiv g_3(\Lambda_c)$ of $g_3$ at some high-energy `cut-off' scale $\Lambda_c$ (which could be the Planck scale,
 or the string scale) by integrating the $\beta$ equation giving the running of $g_3$,
 $ d \ln \mu = d g_3/\beta_3(g_3)$, so that:
\begin{equation}
\label{lambda3}
 \ln \Lambda_3(\Lambda_c,g_c)=  \ln \Lambda_c - \int_{g_*}^{g_c} \frac{d g_3}{\beta_3(g_3)}
\end{equation}
 The expression (\ref{lambda3}) defines $\Lambda_3$ as a function of $\Lambda_c$ and $g_c$.
 If we assume for simplicity that the chosen cut-off $\Lambda_c$ does not depend (in
 the Einstein frame) on $\phi$, the result (\ref{lambda3}) shows that $\Lambda_3$ will inherit
 a $\phi$ dependence from any eventual $\phi$ dependence of $g_c$ according to
 (denoting $\beta_c \equiv \beta_3(g_c)$)
 \begin{equation}
 \frac{\partial \ln \Lambda_3}{\partial \varphi}= - \frac{g_c}{\beta_c}  \frac{\partial \ln g_c}{\partial \varphi}
 \end{equation}
 Similarly, the integration of the RG equation for a running fermionic mass $m_i$, namely
 $d \ln m_i = -d g_3 \gamma_m(g_3)/\beta_3(g_3)$ yields the following expression for
 the value of $m_i$ at the QCD scale, $\ln m_i(\Lambda_3)$:
 \begin{equation}
\label{mlambda3}
 \ln m_i(\Lambda_3)=  \ln m_i(\Lambda_c) + \int_{g_*}^{g_c} \frac{\gamma_m(g_3)}{\beta_3(g_3)} d g_3
\end{equation}
Differentiating this result w.r.t. $\varphi$ then shows that the logarithmic derivative
of $m_i(\Lambda_3)$ w.r.t. $\varphi$ is the sum of two separate contributions, namely
 (denoting $\gamma_c \equiv \gamma_m(g_c)$)
\begin{equation}
 \frac{\partial \ln m_i(\Lambda_3)}{\partial \varphi}= \frac{\partial \ln m_i(\Lambda_c)}{\partial \varphi}
 + \frac{g_c \gamma_c}{\beta_c}  \frac{\partial \ln g_c}{\partial \varphi}
 \end{equation}
On the other hand, by comparing\footnote{In doing this comparison it is useful, as explained above for the
Maxwell action, to provisionally use a `geometric' normalization of the gluon field, i.e. to absorb $g_3$ in $A^A$.}
the $\phi$-dependent terms in (\ref{Lint}) to
the basic action (\ref{action}) (both being considered at the cut-off scale $\Lambda_c$),
we see that the coefficients $d_g$ and $d_{m_i}$ have the effect of adding some $\phi$-dependence
in the values of $g_c$ and $m_i(\Lambda_c)$ of the form
\begin{equation}
\frac{\partial \ln g_c}{\partial \varphi}=- d_g \frac{\beta_c}{g_c} \ , \qquad
 \frac{\partial \ln m_i(\Lambda_c)}{\partial \varphi}= d_{m_i} + \gamma_c d_g.
\end{equation}
 Inserting these results in the $\varphi$-derivatives of $\Lambda_3$ and  $m_i(\Lambda_3)$
 derived above, finally leads (thanks to the cancellation of the $\gamma_c$-dependent
contribution in the derivative of the masses) to the simple results

 \begin{equation}
\frac{\partial \ln \Lambda_3}{\partial \varphi}=  d_g \, ,\qquad
 \frac{\partial \ln m_i(\Lambda_3)}{\partial \varphi}= d_{m_i}.
\end{equation}

 Summarizing: the physical meaning of the five dilaton-coupling coefficients
 $d_a= d_e, d_g, d_{m_e}, d_{m_u}, d_{m_d}$ is (at the linear level in $\phi$) to
 introduce a  $\phi$-dependence in the parameters entering the low-energy physics
 of the form
\begin{eqnarray}
\label{phidependence}
\Lambda_3(\varphi)&=&(1 + d_g \varphi) \Lambda_3, \nonumber \\
\alpha(\varphi) &=   &(1 + d_e \varphi)  \alpha, \nonumber \\
 m_e(\varphi) &=  &(1 + d_{m_e} \varphi) m_e,  \nonumber \\
 \left[m_i(\Lambda_3)\right](\varphi) &=  &(1 + d_{m_i} \varphi) m_i(\Lambda_3), \, i=u,d .
\end{eqnarray}

\subsection{Ratios of dimensional parameters}

Note that a consequence of these equations is that the dimensionless ratios
$m_e/\Lambda_3$, $m_u(\Lambda_3)/\Lambda_3$,  $m_d(\Lambda_3)/\Lambda_3$ depend on
$\varphi$ through the ratios $(1 + d_{m_i} \varphi)/(1 + d_g \varphi) \simeq (1 + (d_{m_i}-d_g) \varphi)$.
In other words, the $\varphi$ sensitivity of these dimensionless ratios is
 \begin{equation}
\frac{\partial \ln \left[m_i(\Lambda_3)/\Lambda_3\right]}{\partial \varphi}= d_{m_i}- d_g \, .
\end{equation}
 Note that this involves only the {\em differences} $d_{m_i}-d_g$. In particular, when the mass couplings $d_{m_i}$
are taken to be all equal to $d_g$, the effect of the $\phi$ couplings is equivalent to introducing
a $\phi$ dependence only in $\Lambda_3$ and $\alpha$. This fact can also be seen by means
of the formulation (\ref{Lintg}) of the $d_g$ coupling. Indeed, when $d_{m_i}=d_g$ the sum of
(\ref{Lintg}) and of the mass-term couplings is equivalent to having a coupling between $\phi$
and the {\em sum} of the anomalous,  $T_{g}^{\rm anom}$, and of the non-anomalous, $T_{g}^{\rm non \, anom}$,
parts of the
trace of the total stress-energy tensor. Therefore, modulo electromagnetic effects, this would imply that
$\phi$ couples to the trace of the total stress-energy tensor, i.e. (using the virial theorem) that
$\phi$ couples to the total mass of the hadron. In this particular case, the only violations of the EP
would come from electromagnetic effects.

However, in view of the fact that the physics which
determines (in the Standard Model) the masses of the leptons and quarks involves the
symmetry breaking of the electroweak sector,
and, in particular, the VEV of the Higgs field, it does not seem a priori likely that a
fundamental theory describing the high-energy couplings of the dilaton can ensure
such a universal feature. From this point of view, one can consider our final results
(\ref{phidependence}) as useful general parametrizations of the low-energy dilaton couplings,
independently of the complicated physics that might connect these parameters to an eventual
high-energy description of the $\phi$ couplings to the fields entering the basic Lagrangian.
For example, heavy quarks do not enter the field couplings (\ref{Lint}), but they enter in the
relation between the QCD scale $\Lambda_3$ (describing the physics at
scales $\lesssim 1$ GeV) and the high-energy boundary conditions, $\Lambda_c, g_c$.
Therefore, the parametrization of $d_g$ in (\ref{phidependence}) implicitly takes into
account the effect of heavy quarks. [Ref. \cite{Kaplan} showed how to explicitly take into account
the effect of heavy quarks, and it is easily checked that their results are in agreement
with the first equation in (\ref{phidependence}).]

We can use the above results to rewrite the expression of the scalar couplings to matter
(\ref{alphaphi}), (\ref{alphavarphi}) in a useful form. As the Planck scale $1/\kappa$ does
not directly enter physics at the QCD scale (besides its possible impact on determining
 $\Lambda_3$ via Eq.(\ref{lambda3})), we can always write the mass of an atom as
\begin{equation}
\label{MA}
m_A = \Lambda_3 M_A(\frac{m_u}{\Lambda_3},\frac{m_d}{\Lambda_3}, \frac{m_e}{\Lambda_3},\alpha ),
\end{equation}
where $M_A$ is a dimensionless quantity, which is a function of the four indicated dimensionless quantities,
say (for later convenience)
\begin{equation}
\label{ka}
(k_u,k_d,k_e,k_{\alpha}) \equiv (\frac{m_u}{\Lambda_3},\frac{m_d}{\Lambda_3}, \frac{m_e}{\Lambda_3},\alpha ).
\end{equation}

Using this notation,  the scalar coupling
to matter  Eq.(\ref{alphavarphi}) can be rewritten (when working in the Einstein frame) as
\begin{equation}
\label{alphaA}
\alpha_A = d_g + \bar{\alpha}_A,
\end{equation}
where $d_g=\frac{\partial \ln \Lambda_3}{\partial \varphi}$ is a universal (non EP-violating) contribution to $\alpha_A$, and where the EP-violating
part $\bar{\alpha}_A$ is given by
\begin{equation}
\label{alphabarAphi}
\bar{\alpha}_A \equiv \frac{\partial \ln M_A}{ \partial \varphi}= \frac{1}{M_A}\frac{\partial M_A}{\partial \varphi}
= \frac{1}{M_A} \sum_{a=u,d,e,\alpha}\frac{\partial M_A}{\partial \ln k_a} \frac{\partial \ln k_a}{\partial \varphi}.
\end{equation}
The logarithmic derivatives of the $k_a$ are given by Eq.  (\ref{phidependence}), so that we can write
more explicitly  $\bar{\alpha}_A$ as the following sum of four contributions:
\begin{equation}
\label{alphabarAdlin}
\bar{\alpha}_A =  \frac{1}{M_A}\frac{\partial M_A}{\partial\varphi}
= \frac{1}{M_A}\left[ \sum_{a=u,d,e} (d_{m_a}-d_g)\frac{\partial M_A}{\partial \ln k_a}+ d_e \frac{\partial M_A}{ \partial \ln \alpha}\right].
\end{equation}

\subsection{Redefining the quark mass parameters}

In the following, we will find it convenient to work with the symmetric and antisymmetric combinations
of the light quark masses, namely
\begin{equation}
\hat{m}=\frac{1}{2}(m_d+m_u) \,, \qquad
\delta m=(m_d-m_u)
\end{equation}
Working in terms of $\hat{m}$ and $\delta m$, means working in terms of mass terms of the form
\begin{equation}
\label{newmassterms}
m_d \bar{d}d +  m_u \bar{u}u   = \hat{m}(\bar{d}d +  \bar{u}u)  + \frac12 \delta m (\bar{d}d -  \bar{u}u)
\end{equation}
which couple to the dilaton as
\begin{equation}
{\cal L}_\phi = .... -\kappa \phi \left[{d}_{\hat m} \hat{m}(\bar{d}d +  \bar{u}u)  + \frac{d_{\delta m}}{2}\delta m (\bar{d}d -  \bar{u}u)\right]
\end{equation}
These definitions are such that, for instance, the coupling of $\varphi$ to $\hat{m}$ is equivalent to
a Hamiltonian coupling of the form,
\begin{equation}
{\cal H} = ....+(1+d_{\hat{m}}\varphi) \hat{m} (\bar{u}u +\bar{d}d ),
\end{equation}
i.e. to introducing a $\varphi$ dependence in the average light quark mass of the type
$ \hat{m}(\varphi) =( 1+d_{\hat{m}}\varphi) {\hat{m}}$.

The link between these new  dilaton-coupling coefficients and the previous ones reads
\begin{equation}
d_{\hat{m}}\equiv \frac{\partial \ln \hat{m}}{\partial\varphi}= \frac{d_{m_d}m_d + d_{m_u}m_u}{m_d+m_u}, \\
d_{\delta m}\equiv \frac{\partial \ln \delta m}{\partial\varphi}= \frac{d_{m_d}m_d - d_{m_u}m_u}{m_d-m_u}.
\end{equation}
In term of this notation  (\ref{alphabarAdlin}) reads
\begin{eqnarray}
\label{alphabarAdlinbis}
\bar{\alpha}_A &= &
 \frac{1}{M_A}\bigl[ (d_{\hat m}-d_g)  \, \hat m \frac{\partial M_A}{\partial {\hat m}} + (d_{\delta m}-d_g) \, \delta m  \frac{\partial M_A}{\partial {\delta m}}   \nonumber \\
 &&+(d_{m_e}-d_g) \, m_e  \frac{\partial M_A}{ \partial_{m_e}} + d_e \, \alpha \frac{\partial M_A}{\partial {\alpha}} \bigl].
\end{eqnarray}

As displayed in  Eq.  (\ref{alphabarAdlinbis}), $\bar{\alpha}_A$ is naturally
decomposed into a sum of four contributions, which are linear in the four dilaton
couplings: $d_{m_a} - d_g$, or $d_e$. Another linear
decomposition can also be applied to the various terms in $\bar{\alpha}_A$: namely the one corresponding to
the various terms in  Eq.  (\ref{massA}). Regrouping some terms in these two possible
linear decompositions, we shall find convenient in our calculations (before coming back
to the more theoretically rooted decomposition (\ref{alphabarAdlinbis})) to decompose $\bar{\alpha}_A$
into three contributions:
\begin{eqnarray}
\label{decompalphabarA}
\bar{\alpha}_A &= &\bar{\alpha}_A^{\rm r \, m \, wo. \, EM}
+  \bar{\alpha}_A^{{\rm bind}} + \bar{\alpha}_A^{d_e}
\end{eqnarray}
where $\bar{\alpha}_A^{\rm r \, m \, wo. \, EM}$ denotes the contribution coming from the terms
linear in the quark and electron masses in the rest-mass contribution (\ref{restmassA}) to $m_A$
({\em without} the electromagnetic contributions), where $\bar{\alpha}_A^{{\rm bind}}$ denotes the contribution coming
from the nuclear binding energy $E^{\rm bind}$ in Eq.  (\ref{massA}), i.e.
\begin{equation}
\label{alphaA3}
\bar\alpha_A^{\rm bind} = \frac{1}{M_A}\frac{\partial (E^{\rm bind}(\varphi)/\Lambda_3) }{\partial \varphi} \, ({\rm with \, fixed \,} \, \alpha ) ,
\end{equation}
and where $\bar{\alpha}_A^{d_e}$
denotes the total electromagnetic contribution, coming both from the EM contributions to the masses
of the nucleons, and from the nuclear Coulomb energy term $E_1$, which is
a part of  $E^{{\rm bind}}$ in Eq.  (\ref{massA}). Note that $\bar{\alpha}_A^{d_e}$ collects the terms
in  $\bar{\alpha}_A$ which are proportional to the EM dilaton coupling $d_e$, i.e. which come from the
$\varphi$ sensitivity of the fine-structure constant $\alpha$. This is why we have
added in the definition of $\bar\alpha_A^{\rm bind}$ above the fact that one must keep $\alpha$
constant when computing it.
As we shall see, the Coulomb energy term plays a special role in that it depends {\em both} on
nuclear-binding effects, and on EM ones. As a consequence it will give two separate
contributions: one to $\bar\alpha_A^{\rm bind}$ and one to $\bar{\alpha}_A^{d_e}$.

\section{Analysis of scalar couplings to the binding energy of nuclei}

We will first focus  on the scalar coupling to the nuclear binding energy, Eq.  (\ref{alphaA3}),
because this term has not yet received a satisfactory treatment in the literature.

When dealing with nuclear binding it is convenient to work with the
(half) sum and difference of the light quark\footnote{As explained above, heavy quarks are assumed to have been integrated out from the theory, thereby producing a shift in the QCD scale $\Lambda_3$,
and its associated dilaton coupling $d_g$. The effect of the strange quark, which is intermediate between
heavy and light, is discussed in the appendix.} masses, $\hat{m}$ and $\delta m$, as
introduced above. Indeed, the quark-mass dependence of nuclear binding is dominated
by its dependence on the average light quark mass $\hat{m}$ because  pion exchanges yield the
dominant contribution to nuclear binding , and pion masses  are proportional to
$\hat{m}$, while they are insensitive to the difference in quark masses. [ The
quark mass difference $\delta m$ is important for the neutron and proton masses,
and will enter the computation below of the rest-mass contribution to EP violation.]

As explained above, the $\varphi$ dependence of  $d_{\hat{m}}$ implies the following result for the `nuclear binding energy'
contribution, Eq.  (\ref{alphaA3}), to EP violation:
\begin{equation}
\bar{\alpha}_A^{{\rm bind}} =
\frac{(d_{\hat{m}}-d_g)}{m_A}\hat{m} \frac{\partial E^{\rm bind} }{\partial \hat{m}}
\end{equation}
In QCD, because the pion is almost a Goldstone boson of the dynamically broken chiral symmetry, the pion mass-squared is linear in the quark mass,
$m^2_\pi \simeq b_0 \Lambda_3 \ \hat{m}$, where $b_0$ is a pure number. This relation is accurate in the physical region, so that we can translate
our formula into one involving the pion mass,
\begin{equation}
\bar{\alpha}_A^{{\rm bind}} = \frac{(d_{\hat{m}}-d_g)}{m_A}m^2_\pi \frac{\partial E^{\rm bind} }{\partial m_\pi^2}
\end{equation}
Our major task then translates into knowing the dependence of nuclear binding on the mass of the pion. 

The semi-empirical mass formula describes the binding energy $m_A - m_A^{\rm rest \, mass}$ through the following terms:
\begin{equation}
 m_A - m_A^{\rm rest \, mass} = E^{\rm bind} ,
\end{equation}
where the nuclear binding energy is approximately described as
\begin{equation}
 E^{\rm bind} =- a_v A +a_s A^{2/3} +a_a \frac{(A-2Z)^2}{A} + a_c \frac{Z(Z-1)}{A^{1/3}}  - \delta \frac{a_p}{A^{1/2}} \, .
\end{equation}

The various contributions to the nuclear binding energy\footnote{Please be aware of a dual
notation in that the letter $A$ is used both as a label for a certain type of atom, and, in the semi-empirical mass formula,
as a notation for the mass number $A=Z+N$.} are called, respectively
the volume energy, the surface energy the asymmetry energy, the Coulomb energy and the pairing energy. [In the latter, $\delta = \frac{1}{2} \, [(-)^N + (-)^Z]$, i.e.
 $\delta=+1$ for even-even nuclei, $\delta=-1 $ for odd-odd nuclei and $\delta=0 $ otherwise.]
Typical fit values for these parameters are \cite{massformula} $a_v =16~ {\rm MeV}, a_s=17 ~{\rm MeV}, a_a= 23 ~{\rm MeV}, a_p= 12 ~{\rm MeV},
a_c = 0.717 ~{\rm MeV} $. Note that, here and in the following, the unit of $1$ MeV is supposed to represent a fixed fraction of the QCD mass scale,
say $\simeq 10^{-3} \Lambda_3$ if we use, as indicated above, a reference value $g_*$ for $g_3$ such that $\Lambda_3 \simeq 1$ GeV.]

The $\hat{m}$ sensitivity of $ E^{\rm bind}$ comes from the $\hat{m} $ sensitivity of the various coefficients $a_v, a_s, a_a, a_c, a_p$ (taken
in units of $\Lambda_3$). We shall discuss successively the $\hat{m}$ sensitivities of: (i) $a_v$ and $a_s$, (ii) $a_a$, and (iii) $a_c$.
Concerning the pairing interaction term $a_p$ we found that it was subdominant in our final results because it is down by a factor of $A^{7/6}$
compared to our primary $A$ dependence. Even when allowing for variations with quark mass comparable to that of the asymmetry
energy we found that it is negligible in the end, so we drop it at this stage.

\subsection{The central nuclear force terms: $a_v$ and $a_s$}

Let us first consider the terms proportional to  $a_v$ and $a_s$. They come from the isospin symmetric central nuclear force,
which is the dominant contribution in the binding of heavy nuclei.
Our previous work \cite{chiral, damourdonoghue} shows that this component has an enhanced dependence on the quark masses and hence it
has an enhanced coupling to a dilaton. This large dependence comes because the central potential involves competing effects of an intermediate
range attractive force and a shorter range repulsive force. The cancelation between these two effects (which are individually of order $ \pm 100$ MeV
per nucleon) lead to a binding energy which is quite small on the QCD scale (namely of order $- 10$ MeV per nucleon).
However, the attractive force is far more sensitive to pion masses because it involves two pion exchange. Changing the pion mass a modest amount
upsets the cancelation of the two components and leads to a larger effect than might naively be expected.

The central force is parameterized by two terms denoting the volume energy and the surface energy,
\begin{equation}
E^{\rm bind} =- a_v A +a_s A^{2/3} + {\rm residual ~terms}
\end{equation}
The central potential is isospin symmetric, and can involve exchanges which carry angular momentum quantum numbers 0 or 1.
The work of Ref.~\cite{furnstahl} uses a general basis of contact interactions \cite{contact} to quantify these
contributions to nuclear binding. This parameterization only assumes that the interactions have a range which is smaller than the momentum in nuclei $k \sim 200$~MeV. The dominant contact interactions are found to be those of an attractive scalar and a repulsive vector, describing the integrated effects of the potentials. They are parameterized by  strengths $G_S ~~(G_V)$ for the scalar (vector) channel. We can then use the results of Ref. ~ \cite{furnstahl} to give the binding energy as a function of these strengths, normalized to
their physical values, by defining
 \begin{equation}
\eta_S \equiv \frac{G_S}{G_S|_{\rm physical}},
~~~~~~
\eta_V \equiv \frac{G_V}{G_V|_{\rm physical}}
\end{equation}
This results in
\begin{equation}
E^{\rm bind} = -(120 A -{97}{A^{2/3}}) \eta_S + (67A -
{57}{A^{2/3}} )\eta_V  +{\rm residual ~terms}
\end{equation}
where the numbers are in units of MeV. One can see here the cancelation between the primary terms as each is larger than their sum.
Of these two contributions, our calculations indicate that it is the {\em scalar channel} ($\eta_S$) that has the most important effect.
This is because the scalar channel is dominated by the exchange of two pions, which is highly sensitive to the pion mass. While
the two pion contribution is often parameterized by an effective sigma meson, the low energy exchange of two pions is required in chiral
perturbation theory and is calculable\footnote{Other estimates of mass dependence \cite{twonucleon} have not explicitly taken into account
this low energy effect.}. This accounts for much of the strength typically ascribed to the sigma \cite{sigma}. The vector interaction has
a very small low energy contribution from three pions, and estimates of the quark mass dependence of the mass and couplings of a massive
vector boson indicate a tiny residual contribution \cite{chiral}.

\begin{figure}[ht]
 \begin{center}
  \includegraphics[scale=1.0]{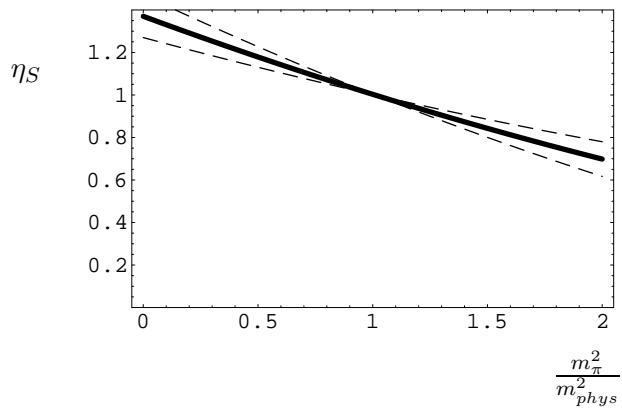}
 \end{center}
 \caption{\small{The value of the scalar strength $\eta_S$ as a function of the pion mass. }}
 \label{spectral}
\end{figure}

With these results we have argued that the main contribution is the variation of the scalar strength with quark mass,
 \begin{equation}
\bar{\alpha}_A^{{\rm bind}} = -\frac{(d_{\hat{m}} -d_g)}{m_A}(120 A -{97}{A^{2/3}})
m^2_\pi \frac{\partial \eta_S}{\partial m_\pi^2}
\end{equation}
We use the result of Ref.  \cite{chiral}, displayed in Fig \ref{spectral} showing the scalar strength as a function of the pion mass. This variation arose almost entirely from the threshold modification in the two pion effects at low energy, where the chiral techniques are most reliable and where we expect the greatest sensitivity to a change in the mass \cite{eft,dispersive}.
We can use this directly to obtain
\begin{equation}
\label{etaS}
\hat{m} \frac{\partial \eta_S }{\partial \hat{m}}=m^2_\pi \frac{\partial \eta_S}{\partial m_\pi^2} =-0.35\pm 0.10
\end{equation}
The error bar comes from uncertainties in the chiral expansion. We will not display the error bar in subsequent formulas, but all results in the binding energy carry this level of uncertainty. Our final result for the central dependence in the dilaton coupling is
\begin{eqnarray}
\label{alphacentral}
\bar{\alpha}_A^{{\rm bind}}|_{\rm central} &=& \frac{(d_{\hat{m}} -d_g) }{m_A}(42 A -{34}{A^{2/3}}) ~~({\rm MeV})\nonumber \\
&\approx&(d_{\hat{m}} -d_g)~F_A~\left( 0.045 -\frac{0.036}{A^{1/3}}\right)~.
\end{eqnarray}
In the final line we have introduced the notation
 \begin{equation}
\label{FA}
F_A \equiv \frac{A \, m_{\rm amu}}{m_A}
\end{equation}
where $m_{\rm amu}=931$~MeV is the atomic mass unit
(i.e. the nucleon mass $m_N=939$~MeV minus the
average binding energy per nucleon, $\simeq 8$ MeV). The
factor $F= Am_{\rm amu}/m_A$ remains quite close to one all over the periodic table (modulo $O(10^{-3})$). Note that our result Eq. (\ref{alphacentral}) for the light-quark-mass ($\hat{m}$) dependence
is significantly larger (by a factor $2.2$) than the estimate used by Dent \cite{dent}. Indeed,
Eq. (\ref{alphacentral}) corresponds, say for the crucial surface energy, to a logarithmic
sensitivity $\partial \ln a_s/ \partial \ln \hat{m} =- 34 \,{\rm MeV}/a_s=-2$, while Ref. \cite{dent}
estimated $\partial \ln a_s/ \partial \ln \hat{m} \simeq - 0.9$.

\subsection{The asymmetry energy term: $a_a$}

Let us now discuss the $\varphi$ sensitivity of the asymmetry energy $\propto a_a$ which is, after the volume,
 surface and Coulomb terms, the fourth dominant contribution to $E^{\rm bind}$.
The asymmetry energy has two components. The first comes from the Pauli principle which requires that, when there is
an excess of neutrons over protons, the extra neutrons must be placed into higher energy states than the protons. The other
component is due to the nuclear force in which the isospin dependent interactions create a stronger attraction for an neutron and
proton compared to two neutrons or two protons.

The asymmetry energy has been calculated by Serot and Walecka \cite{serot} in the same framework that we use in our work
on nuclear matter  \cite{damourdonoghue}. This takes the form
\begin{equation}
\label{aa}
a_a =  \frac{k^2_F}{6\sqrt{M_*^2 + k_F^2}}+ \frac{G_\rho}{12\pi^2}k_F^3
\end{equation}
where
\begin{equation}
M_* = m_N \left(1+ \frac{\gamma G_Sk_F^3}{6\pi^2}\right)
\end{equation}
is the nucleon mass modified by interactions in nuclear matter (with $G_S < 0$ so that $M_* < m_N$). For isoscalar nuclear matter we have $\gamma = 4$.
In meson exchange models $G_\rho = g^2_\rho/m^2_\rho$ and $G_S = - g^2_\sigma/m^2_\sigma$ are the vector meson and scalar coupling
strengths. The $k_F^3$ dependence in the second term in $a_a$ comes from a calculation of the nuclear density in terms of the Fermi momentum $k_F$.

As mentioned above, our estimates indicate that the mass dependence of the vector meson coupling strength is weak. However, the Fermi
 momentum depends on the scalar strength, which has a sizeable mass variation. The Fermi momentum increases as the scalar strength increases.
We calculate this through our work on nuclear matter in which we solve for the Fermi momentum as a function of the scalar
strength (e.g. see Fig. 4 of  \cite{damourdonoghue}). More precisely, using our approximate
analytical model, Eq. (17) of \cite{damourdonoghue}, with the values $G_S=-355.388\,$GeV$^{-2}$,
and $G_S=+ 262.89\,$GeV$^{-2}$ (which entail the phenomenologically good values $a_v=15.75\,$MeV
and $k_F=1.30\,$fm$^{-1}$) we find that
\begin{equation}
\frac{\partial \ln k_F}{\partial \ln G_S} \simeq 0.525
\end{equation}
Using that dependence we find that both components of the asymmetry energy
in Eq. (\ref{aa})  vary in the same direction with
the scalar strength. The kinetic contribution (first term) varies with a logarithmic rate $\simeq 2.54$,
while the other one varies like $k_F^3$, i.e. with a logarithmic rate $3 \times 0.525 = 1.575$.
The combination of the two contributions then varies with a rate
\begin{equation}
\frac{\partial \ln a_a}{\partial \ln G_S} \simeq 2.35
\end{equation}
Combining this variation with the logarithmic mass variation Eq. (\ref{etaS}) of the scalar strength $G_S$ then yields
\begin{equation}
m_\pi^2 \frac{\partial a_a}{\partial m_\pi^2} =  \frac{\partial a_a}{\partial G_S} m_\pi^2 \frac{\partial G_S}{\partial m_\pi^2} =-0.82 \, a_a = -19 ~{\rm MeV}
\end{equation}
Note that our framework shows that $\partial \ln a_a/ \partial \ln \hat{m} \simeq - 0.82$
is rather different from $\partial \ln a_s/ \partial \ln \hat{m} = -2$. This shows again
the subtlety of quark-mass effects in nuclear physics.

\subsection{The Coulomb energy term: $a_c$}

The Coulomb energy also has a dependence on the strong interaction coupling terms. Dimensionally this is because the electromagnetic coupling $\alpha$ is dimensionless, so that the overall energy scale associated with $a_c$ comes from the nuclear interactions. Physically, this dependence is also logical because the Coulomb energy depends on how tightly the nucleons are packed together. We estimate this effect in this subsection.

An approximate analytic expression for the coefficient of the Coulomb contribution to the
nuclear binding energy is $ a_c \simeq (3/5) \alpha/r_0$
where $r_0 \simeq 1.2$ fm is the scaled nuclear radius: $r_A = r_0 A^{1/3}$. Writing that the total baryonic number
within the volume of the nucleus, i.e. $\rho_B 4 \pi r_A^3/3$ (with $\rho_B = \gamma k_F^3/(6 \pi^2)$) is equal to $A$,
one gets the link $ k_F r_0 =( 9 \pi/8)^{1/3}$. Therefore, $r_0$ varies inversely proportionally to $k_F$, so that
the above result shows that $a_c \propto \alpha k_F$. This yields a logarithmic sensitivity of $a_c$ to
variations of $G_S$ with the same rate as $k_F$ itself, i.e. $0.525$, as quoted above. Multiplying this rate
by the rate $-0.35$ of Eq. (\ref{etaS}), then yields
\begin{equation}
\hat{m} \frac{\partial E_1 }{\partial \hat{m}} = - 0.184 a_c \frac{Z(Z-1)}{A^{1/3}} = - 0.13 \frac{Z(Z-1)}{A^{1/3}} {\rm MeV}
\end{equation}

\subsection{The complete scalar coupling to the binding energy}

Combining our partial results, we finally obtain for $\bar{\alpha}_A^{{\rm bind}}$
the following sum
\begin{eqnarray}
\label{baralphabind}
\bar{\alpha}_A^{{\rm bind}} &= &(d_{\hat{m}} -d_g)  F_A  \\
&\times&\left[ 0.045 -\frac{0.036}{A^{1/3}} - 0.020 \frac{(A-2Z)^2}{A^2}  - 1.42 \times 10^{-4} \, \frac{Z(Z-1)}{A^{4/3}} \right]. \nonumber
\end{eqnarray}
In writing this result, we have, as above, factorized $F_A =A m_{\rm amu}/m_A$.

\section{Scalar couplings to the rest mass of atoms}

In this section we study the first term on the r.h.s. of Eq.~(\ref{decompalphabarA}), i.e. the contribution to $\bar\alpha_A$ coming from
the $\varphi$ sensitivity of the rest masses of the low-energy constituents of atoms, namely protons, neutrons and electrons (\`a la \cite{Damour:1994ya}).

In view of the expression Eq.~(\ref{newmassterms}) for the mass terms of the light quarks, we can write the masses of the nucleons as \cite{quark masses}
\begin{eqnarray}
\label{eq4.1}
m_p &= &m_{N3} + \sigma - \frac{1}{2} \, \delta + C_p \, \alpha \, , \nonumber \\
m_n &= &m_{N3} + \sigma + \frac{1}{2} \, \delta + C_n \, \alpha \, ,
\end{eqnarray}
where $m_{N3}$ is the nucleon mass in the ``chiral limit'' of massless light\footnote{In the present treatment, we absorb in
$m_{N3} \propto \Lambda_3$ the EP non violating effect of the strange quark; see the appendix.} quarks, and where the electromagnetic
contributions $C_p \, \alpha$, $C_n \, \alpha$ will be ignored here and treated in the next section. The quantities $\sigma$ and $\delta$
in Eq.~(\ref{eq4.1}) denote the matrix elements of the isoscalar $(\propto \bar d d + \bar u u)$ and isovector $(\propto \bar d d - \bar u u)$
terms in a neutron state:
\begin{eqnarray}
\label{eq4.2}
\sigma &=& \langle n| \hat{m} (\bar{d}d+\bar{u}u)|n \rangle \nonumber\\
\delta &=& \langle n| (m_d-m_u) (\bar{d}d-\bar{u}u)|n \rangle
\end{eqnarray}
These combinations of the  quark mass contributions to the individual nucleons are reasonably well known.
The isoscalar contribution is related to the $\pi N$ sigma term and has the
value $\sigma = 45$~MeV \cite{Gasser:1990ce}. The isovector difference can be obtained by SU(3) sum rules
\begin{equation}
\delta = \frac{m_d-m_u}{m_s-\hat{m}}[m_\Xi -m_\sigma] = 3.1~ {\rm MeV}
\end{equation}
The $\varphi$ sensitivity of the rest mass contribution (without the EM contribution) of an atom,
\begin{equation}
\label{eq4.3}
m_A^{\rm r \, m \, wo. \, EM} = A \, m_{N_3} + A \, \sigma + \frac{1}{2} \, (N-Z) \, \delta + Z \, m_e \, ,
\end{equation}
comes from the fact that $\sigma \propto \hat m (\varphi)$, $\delta \propto \delta m(\varphi)$, and
from the $\varphi$ dependence of $m_e$. Using our general results above, we therefore have
\begin{eqnarray}
\label{eq4.4}
\bar\alpha_A^{\rm r \, m \, wo. \, EM} &= &(d_{\hat m} - d_g) \, \frac{A \sigma}{m_A} + \frac{1}{2} (d_{\delta_m} - d_g) \, \frac{(N-Z) \delta}{m_A} \nonumber \\
&+ &(d_{m_e} - d_g) \, \frac{Z m_e}{m_A} \, .
\end{eqnarray}
Inserting the numerical values of $\sigma , \delta$ and $m_e$ yields
\begin{eqnarray}
\label{eq4.5}
\bar\alpha_A^{\rm r \, m \, wo. \, EM} &\simeq & F_A\Bigl[ 0.048 (d_{\hat m} - d_g) + 0.0017 (d_{\delta_m} - d_g) \, \frac{A-2Z}{A} \nonumber \\
&+ &5.5 \times 10^{-4} (d_{m_e} - d_g) \, \frac{Z}{A} \Bigl]  \, .
\end{eqnarray}

\bigskip


\section{Electromagnetic effects}

In this section, we review the electromagnetic coupling, which is contained in the Lagrangian,
i.e. the contribution
\begin{equation}
\alpha_A^{(d_e)}  = \frac{d_e}{m_A}\alpha \frac{\partial m_A }{\partial \alpha}
\end{equation}

The main electromagnetic effects in the atomic masses come from the electromagnetic shifts in the nucleon masses and from the
electromagnetic contribution to nuclear binding, $E_1$.
\begin{equation}
\alpha_A^{(d_e)} =  \frac{d_e}{m_A}\left[Z \alpha \frac{\partial m_p }{\partial \alpha} + (A-Z) \alpha \frac{\partial m_n }{\partial \alpha} +\alpha \frac{\partial E_1 }{\partial \alpha}               \right]
\end{equation}
We follow Gasser and Leutwyler \cite{quark masses} in the estimate of the electromagnetic portions of the proton and neutron masses
\begin{equation}
\alpha \frac{\partial m_p }{\partial \alpha} = C_p=0.63~ {\rm MeV} ~~~~~~~~\alpha \frac{\partial m_n }{\partial \alpha}=C_n==-0.13 ~{\rm MeV}
\end{equation}
The electromagnetic binding is known from the semi-empirical mass formula
\begin{equation}
\alpha \frac{\partial E_1 }{\partial \alpha}  =a_c\frac{Z(Z-1)}{A^{1/3}}
\end{equation}
with $ a_c = 0.717 ~{\rm MeV} $. These combine to yield
\begin{equation}
\label{baralphade}
\bar{\alpha}_A^{(d_e)} = d_e  F_A \left[ -1.4 + 8.2 \frac{Z}{A} + 7.7 \frac{Z(Z-1)}{A^{4/3}}  \right]\times 10^{-4}
\end{equation}
As above, the factor $F_A= A m_{\rm amu}/m_A$ can be replaced by one in lowest approximation.

\section{Implications for the Equivalence Principle }

\subsection{General parameterization}

Summarizing our results, the dilaton coupling to an atom can be written as
\begin{equation}
{\alpha}_A = d_g +  \bar{\alpha}_A^{\rm r \, m \, wo. \, EM} + \bar{\alpha}_A^{{\rm bind}} + \bar{\alpha}_A^{(d_e)}
\end{equation}
where $\bar{\alpha}_A^{\rm r \, m \, wo. \, EM}$ is given by  Eq. (\ref{eq4.5}), $\bar{\alpha}_A^{{\rm bind}}$ by Eq. (\ref{baralphabind}),
and $ \bar{\alpha}_A^{(d_e)}$ by  Eq. (\ref{baralphade}). It will be convenient for the following to rewrite this result as
\begin{equation}
\label{alpha5}
{\alpha}_A = d_g +  \bar{\alpha}_A
\end{equation}
with the decomposition
\begin{equation}
\label{baralpha4}
\bar{\alpha}_A =  \left[ (d_{\hat m} - d_g) Q_{\hat m} + (d_{\delta m} -d_g) Q_{\delta m} + (d_{m_e} - d_g) Q_{m_e} + d_e Q_e \right]_A
\end{equation}
where $Q_{k_a}$ can be thought of as the ` dilaton charge' coupled to the parameter $k_a$. These are given by
\begin{equation}
\label{Qmhat}
 Q_{\hat m} = F_A \left[ 0.093 -\frac{0.036}{A^{1/3}} - 0.020 \frac{(A-2Z)^2}{A^2}
- 1.4 \times 10^{-4} \, \frac{Z(Z-1)}{A^{4/3}} \right]  ,
\end{equation}
\begin{equation}
\label{Qdeltam}
Q_{\delta m} = F_A \left[0.0017   \, \frac{A-2Z}{A} \right] ,
\end{equation}
\begin{equation}
\label{Qme}
 Q_{m_e} = F_A \left[ 5.5 \times 10^{-4}  \, \frac{Z}{A} \right] ,
\end{equation}
and
\begin{equation}
\label{Qe}
Q_e = F_A   \left[  -1.4 + 8.2 \frac{Z}{A} + 7.7 \frac{Z(Z-1)}{A^{4/3}}  \right]\times 10^{-4}.
\end{equation}
Here, as above, the factor $F_A$ denotes $F_A \equiv A m_{\rm amu}/m_A$ (it can be replaced by one in lowest approximation).

\subsection{Relation to theoretical expectations}

Note that all the various contributions to the non-universal part $\bar{\alpha}_A$ of $\alpha_A = d_g + \bar{\alpha}_A$ contain small
numerical coefficients in front of the various basic dilaton couplings $d_g, d_e, d_{\hat{m}},d_{\delta m}, d_{m_e}$. It is therefore
a priori probable that the composition-dependent part $\bar{\alpha}_A$ is small compared to the composition-independent\footnote{
Actually, if we define the composition-independent part of  $\alpha_A$ by some average over the composition of the bodies relevant
for the considered gravity tests,  $\alpha_A^{\rm c.i.}$ will have, besides $d_g$, a contribution coming from $\bar{\alpha}_A$,
and notably from terms $\sim 0.1 (d_{\hat{m}}-d_g)$ coming fron the QCD binding of nucleons, and the nuclear
binding of nuclei. To simplify our discussion we shall assume that these terms are small. It is enough to replace some
of our factors $d_g$ below by $\alpha_A^{\rm c.i.}= d_g^{*}\simeq d_g + 0.1 (d_{\hat{m}}-d_g) + \cdots$ to refine our estimates.}
part $\alpha_A^{\rm c.i.} = d_g$.

We recall that the latter composition-independent part is, in principle, accessible in various experimental tests of relativistic gravity.
For instance, in the notation of tests of post-Newtonian gravity, $\alpha_A^{\rm c.i.} = d_g$, is related to
the Eddington parameter $\gamma$ via (see, e.g., \cite{Damour:1992we})
\begin{equation}
 \gamma -1 = - 2 \frac{d_g^2}{1 + d_g^2} \simeq - 2 d_g^2
\end{equation}
The most precise current test of relativistic gravity \cite{Bertotti:2003rm} constrain $(\gamma -1)/2$, i.e. $d_g^2$ at the level
\begin{equation}
\label{cassini}
 d_g^2\simeq \frac{1-\gamma}{2} < 10^{-5}
\end{equation}
Planned improved solar-system tests might improve this limit to the $ 10^{-7}$ level. As we are going to see, and
as was pointed out by many authors before (see, e.g.  \cite{Damour:1995gi}), such levels are much less constraining than
the ones accessible by experimental tests of the EP.

By contrast to the composition-independent tests whose signals are proportional to  $d_g^2$, the EP-violation signals
will all be (see Eq. (\ref{da/a})) proportional to
\begin{equation}
\alpha_E (\alpha_A- \alpha_B) \simeq \alpha^{\rm c.i.}(\bar{\alpha}_A- \bar{\alpha}_B)
\end{equation}
Therefore EP signals will involve the  product of $d_g$ (or rather $d_g^{*} =\alpha^{\rm c.i.}$)
by one of the other dilaton couplings
entering the $\bar{\alpha}_A$'s, i.e. they will be proportional to a combination of terms
involving the following four coefficients
\begin{equation}
\label{EPparameters}
 d_g^* \, (d_{\hat{m}}-d_g), d_g^* \, (d_{\delta m}-d_g ), d_g^* \, (d_{ m_e}-d_g ) \quad  {\rm or} \quad d_g^* \, d_e
\end{equation}

This raises several issues of direct phenomenological interest:
(i) can, in principle, EP experiments measure all four (a priori independent) parameters (\ref{EPparameters}) ?;
(ii) are there theoretical arguments suggesting that, among all the EP signals associated to these parameters,
some of them might dominate over the others?

Concerning the first question (which has also been addressed in \cite{dent}),
let us note that if we approximate the factor $F_A =A m_{\rm amu}/m_A$ by one ( and  $Z(Z-1)$ by  $Z^2$) ,
 the composition dependence of our general dilaton coupling above will vary, along the periodic table, according to
\begin{equation}
\label{compdependence}
\bar{\alpha}_A = a_0 + \frac{a_1}{A^{1/3}} + a_2 \frac{A-2Z}{A} + a_3\frac{(A-2Z)^2}{A^2} + a_4\frac{Z^2}{A^{4/3}}
\end{equation}
where the five coefficients $a_0, \ldots, a_4$ are linear combinations (which are easily read off the results above) of the four
dimensionless dilaton couplings $ d_{\hat{m}}-d_g,d_{\delta m}-d_g, d_{m_e}-d_g, d_e$. Here the constant offset $a_0$ is not
measurable\footnote{At least in our approximation  $A m_{\rm amu}/m_A \simeq 1$. If one were to keep the small fractional
($\sim 10^{-3}$) variations of the ratio $A m_{\rm amu}/m_A$, one might measure part of the $a_0$ coefficient.}
in EP experiments. By contrast, EP experiments can, in principle, measure the coefficients of the four
different composition-dependences associated with $a_1, a_2, a_3, a_4$. Barring some degeneracies, this means
that, in principle, a well-devised set of ideal EP experiments could measure the four theoretical
parameters (\ref{EPparameters}) [see, e.g., \cite{damour} for discussions of the related optimization of
the choice of materials in EP experiments, and \cite{dent} for an example of the determination of four
theoretical parameters from four independent EP data].

However, EP experiments will be more likely to detect signals
associated with functions of $A$ and $Z$ that vary significantly over the periodic table. From this point of view,
two signals, among the four ones in Eq. (\ref{compdependence}), are likely to be more prominent: namely the ones
associated to the parameters $a_1$ and $a_4$. Indeed, both $A^{-1/3}$ and $Z^2A^{-4/3}$ vary significantly
along the periodic table. By contrast, the quantities $(A-2Z)/A$ and $ ((A-2Z)/A)^2$ vary only mildly.
Indeed, the `valley' of stable nuclei is located along a specific line in the $A,Z$ plane which is rather
close to the $A= 2 Z$ (i.e. $N=Z$) straight line. Actually, in absence of the Coulomb repulsion between
protons, the Pauli principle would favour an equal number of protons and neutrons (cf. the discussion
of the asymmetry energy above). The Coulomb effects modify this in favouring a relatively small excess of
neutrons over protons. More precisely, the bottom of the valley of stable nuclei is
around \cite{massformula}
\begin{equation}
\label{Z}
 Z_{\rm stable} \simeq \frac{1}{2} \frac{A}{1 + 0.015 A^{2/3}}
\end{equation}
Using this result we see that $(2Z-A)/A \simeq (1+0.015 A^{2/3})^{-1} -1$, which is small and whose variation with $A$ is reduced by the small coefficient $0.015$.

In conclusion, the two EP signals that are probably most easily measurable in  Eq. (\ref{compdependence})
are the ones associated to  $A^{-1/3}$ and $Z^2A^{-4/3}$.
In previous work on the phenomenological consequences of dilaton couplings  \cite{damour,Damour:1994ya} it was
suggested that the EP signal would be essentially proportional to $Z^2A^{-4/3}$, i.e. related to
the Coulomb energy term $\propto d_e$ in the results above. Our analysis of the quark-mass sensitivity
of nuclear binding is now modifying this conclusion in suggesting that the $\varphi$ dependence of atomic masses
will contain, in addition to this Coulomb-related term, another term (related to the quark-mass dependence
of nuclear binding), with a  $A^{-1/3}$ variation over the periodic table.

An important issue is to know whether theoretical considerations can tell us a priori
something about the relative order of magnitude of these Coulomb and nuclear terms. In order
to discuss this we need to know something about the expected  relative magnitude of
 $ d_g^* d_e$ versus $d_g^* (d_g- d_{\hat{m}})$, i.e. the relative magnitude of $d_e$ versus $d_g- d_{\hat{m}}$.
We shall next argue that it is theoretically plausible either that $d_e \sim d_g- d_{\hat{m}}$,
or that $d_e \sim ( d_g- d_{\hat{m}})/40$.

Indeed, we have seen above that our dilaton coefficients $d_g, d_{\hat{m}}, d_e $ where respectively defined as being the logarithmic derivatives
of $\Lambda_3, \hat{m}, \alpha$. On the other hand, it is natural to consider (at least in string theory) that a dilaton couples
with roughly equal strengths to the various terms in the Wilsonian action considered at some high-energy `cut-off'
scale $\Lambda_c$, near the  {\em string scale},  i.e. probably near
 the Planck scale $ m_P= 1/\kappa \sim 3.44 \times 10^{18}$ GeV. If this is the case, the relative magnitudes of the low-energy
dilaton couplings $d_g, d_{\hat{m}}, d_e $ is determined by the functional dependences that relate the low-energy quantities
$\Lambda_3, \hat{m}, \alpha$ to basic couplings at the string, or Planck, scale. In the case, of the fine-structure constant,
though it does run, according to the RG, between the IR (i.e. $m_e$) and the GUT or Planck scale, this running is relatively small because of
the smallness of the factor $(2 \alpha/3 \pi)$ which multiplies $\ln (m_P/m_e)$. As a consequence, one expects that the low-energy
EM dilaton coupling $d_e$ is similar to its more fundamental high-energy counterpart. [This is also related to the fact that
we could neglect, in our action (\ref{Lint}) the EM analog of the ratio $\beta_3(g_3)/g_3$
(i.e. $\beta_3(g_3)/g_3^3$ with geometrically normalized gauge fields), because
 $\beta_{EM} (e)/e^3$ is essentially constant.] The situation is, however, quite different for the low-energy coupling $d_g$ to the gluon field
energy. There are two equivalent ways of seeing it. One way (used in \cite{Kaplan}) precisely consists in drawing the consequences
of having a factor $\beta_3(g_3)/g_3$ in front of $(F^A)^2$ (to ensure RG invariance). When comparing the matching of this factor at the Planck scale,
versus its meaning at the low-scale $\Lambda_3 \sim 1$ GeV, one sees that $d_g$ differs from its high-energy counterpart by a largish factor of order
\begin{equation}
 K = f_{\rm h.q.} \, \frac{g_3(\Lambda_c)}{\beta_3(\Lambda_c)}
\end{equation}
where the additional  factor $f_{\rm h.q.}$ takes into account the effect of the heavy quarks \cite{Kaplan}. The second way
(used in \cite{Damour:1994ya}) consists in
differentiating the expression giving $\Lambda_3$ in terms of high-energy boundary conditions. We have seen above that the
definition of $\Lambda_3$ coming from the integration of the RG-running equation for $g_3$ yields equivalent results,
with the same appearance of the largish factor  $g_3(\Lambda_c) / \beta_3(\Lambda_c)$. It is easily checked that
this second way also automatically includes the effect of heavy quarks, i.e. the factor $f_{\rm h.q.}$ in $K$.
Actually, this second way provides a quick way to estimate the order of magnitude of the factor $K$ above. Indeed,
the reason why $\Lambda_3$ is herarchically smaller than $\Lambda_c$ is that solving the RG-running equation leads
to a result of the type $\Lambda_3 \sim \Lambda_c \exp (- C/g_c^2)$.  Differentiating this expression w.r.t. $\varphi$
immediately shows that the amplification factor between $d_g$ and the high-energy dilaton coupling
$\partial \ln g_c^2/ \partial \varphi$
can be written as
\begin{equation}
 K = \ln( \Lambda_c/\Lambda_3)
\end{equation}
Using, for instance, $\Lambda_c \sim m_P= 1/\kappa \sim 3.44 \times 10^{18}$ GeV then yields
$K \sim \ln (m_P/1 {\rm GeV}) \sim 42.7$, as in Ref. \cite{Damour:1994ya}, and consistently with the results of
\cite{Kaplan}, for the MSSM case. [We note also that the presence of this logarithmic enhancement factor in
the dilaton coupling was pointed out in Ref. \cite{Taylor:1988nw}.]

When considering the low-energy dilaton coupling to the average light quark mass  $\hat{m}$, the second way of computing it
similarly suggests that it will contain a large enhancement factor $\sim \ln( \Lambda_c/\hat{m})$ with respect to some
high-energy counterpart that should a priori be comparable  to $\partial \ln g_c^2/ \partial \varphi$. Indeed, let us recall
that the quark masses are of order $m_q \sim f H$, where $H$ is the Higgs's VEV, and $f$ a dimensionless Yukawa coupling. As we do not know
what is the mechanism which determines (from the UV) the scale of the electroweak breaking (i.e. which allows for a negative squared mass for the Higgs
at low energies), we cannot compute the sensitivity of $m_q$ to $\varphi$. However, it is plausible, as indicated by the `no-scale'
models \cite{Ellis:1984bm}, that $H$ is related to $\Lambda_c$, via the RG-running of (scalar) masses, by an exponential factor
similar to the one linking  $\Lambda_3$ to $ \Lambda_c$: more precisely, in these models
one has $H \sim \exp(-C'/h_t^2)$, where $C'$ is a constant of order unity, and where $h_t$ is
the Yukawa coupling of the top quark.
 Then, the $\varphi$-derivative of $\ln m_q$ will also contain an
enhancement factor of order $\ln (\Lambda_c/\Lambda_3)$, i.e. of the same order as the enhancement $K$ above,
but probably differing by a factor of order unity.

Summarizing: it seems theoretically plausible that, starting from dilaton couplings which are of the same order, say $d_c = \partial \ln g_c^2 / \partial \varphi$,
when considered at the high-energy scale $\Lambda_c$, the low-energy coupling EM $d_e$ will remain
$d_e \sim d_c$, while $d_g$ and the various $d_{m_a}$ will be enhanced by factors of order $K_a \sim \ln (\Lambda_c/m_a) \sim 40$.
Notably, we can expect $d_g \sim K d_c$, and  $d_{\hat{m}} \sim K' d_c$. This leaves us with the problem of
estimating the difference $d_g - d_{\hat{m}}$ which enters in composition-dependent effects. It is formally of order
$ \sim (K - K') d_c$. We do not know to what extent there could be a compensation between $K$ and $K'$. If such
a compensation exists, i.e. if $K - K' \sim 1$, instead of $\sim 40$, one will have
$ d_g - d_{\hat{m}} \sim d_c \sim d_e$. On the other hand, if $K$ and $K'$ differ by a factor of order unity (or have a different sign),
we will have $d_g - d_{\hat{m}} \sim  40 d_c \gg d_e$. Therefore, we can only write an approximate
link of the type $d_e \lesssim d_g- d_{\hat{m}}$. For our discussion of the relative importance of various
EP signals, it would be too restrictive to assume that Nature has chosen the case where $d_e$ is significantly
smaller than $d_g- d_{\hat{m}}$. We shall therefore continue our discussion under the general assumption
$d_e \sim d_g- d_{\hat{m}}$.

\subsection{Simplified parameterization}

Our theoretical treatment of nuclear binding effects has given us some specific predictions
for the numerical coefficients of the various contributions to the `dilaton charges' $Q_{k_a}$.
To better delineate what they imply for the phenomenology of EP experiments we shall henceforth
make some further approximations. First, we replace the overall factor $F_A = A m_{\rm amu}/m_A$ by one.
This is allowed because we shall see that the leading terms in the $Q_{k_a}$'s vary by factor
of a few over the periodic table, while $F_A$ differs from one only at the $10^{-3}$ level.
The second approximation consists in using the approximate equation (\ref{Z}) to estimate
various $Z$-dependent terms in the dilaton charges. Namely, using this link,
and taking into account the predicted numerical coefficients in the dilaton charges,
one finds that the terms $0.020 (A-2Z)^2/A^2$ (in $ Q_{\hat m}$), and $0.0017 (A-2Z)/A$ (in $Q_{\delta m} $),
are numerically subdominant.[We assume here that, e.g., $d_{\delta m} -d_g  \sim  d_{\hat m} -d_g$ etc.]
In addition, we find that we can replace $Z/A$ by $1/2$ in $Q_{m_e}$ and $Q_e$. After these simplifications,
we can move some left-over composition-independent numerical coefficients out of the $Q$'s, and into the
general composition-independent contribution $d_g$ in ${\alpha}_A$.

After these approximations, we end up with
\begin{equation}
\label{approxalphaA}
{\alpha}_A \simeq d_g^* + \left[ (d_{\hat m} - d_g) Q'_{\hat m}  + d_e Q'_e \right]_A
\end{equation}
where
\begin{equation}
 d_g^* = d_g + 0.093 (d_{\hat m} - d_g) + 0.00027 d_e
\end{equation}
and where
\begin{equation}
 Q'_{\hat m} = -\frac{0.036}{A^{1/3}} - 1.4 \times 10^{-4} \, \frac{Z(Z-1)}{A^{4/3}}
\end{equation}
and
\begin{equation}
 Q'_{e} =  + 7.7 \times 10^{-4} \frac{Z(Z-1)}{A^{4/3}} .
\end{equation}
We think that these approximate expressions capture all the potentially dominant EP violation effects.
We illustrate the variation of these approximate dilaton charges over the periodic table
by giving in Table 1 their values for a sample of elements. [Our table considers many of the
same elements as Table 1 of \cite{dent}, but the crucial new information we provide are
the numerical factors in the charges, as predicted from our results. We use the (non-integer) atomic
weights as an approximate way of averaging\footnote{Essentially we are using the
approximation $\langle f(A) \rangle \simeq f(\langle A \rangle)$,
which is valid to first order for a smooth function $f(A)$.} the result over the natural isotopic composition. ]

\begin{table}[h]\centering
\caption{Approximate EP-violating `dilaton charges' for a sample of materials. These charges are averaged over the (isotopic or chemical, for SiO$_2$) composition.}
\begin{tabular}{ccccc}
\\
{\rm Material} &$A$ &$Z$ &$-Q'_{\hat m}$ &$Q'_e$ \\ \\
{\rm Li} &7 &3 &18.88 $\times 10^{-3}$ &0.345 $\times 10^{-3}$ \\
{\rm Be} &9 &4 &17.40 $\times 10^{-3}$ &0.494 $\times 10^{-3}$ \\
{\rm Al} &27 &13 &12.27 $\times 10^{-3}$ &1.48 $\times 10^{-3}$ \\
{\rm Si} &28.1 &14 &12.1 $\times 10^{-3}$ &1.64 $\times 10^{-3}$ \\
{\rm SiO$_2$} &... &... &13.39 $\times 10^{-3}$ &1.34 $\times 10^{-3}$ \\
{\rm Ti} &47.9 &22 &10.28 $\times 10^{-3}$ &2.04 $\times 10^{-3}$ \\
{\rm Fe} &56 &26 &9.83 $\times 10^{-3}$ &2.34 $\times 10^{-3}$ \\
{\rm Cu} &63.6 &29 &9.47 $\times 10^{-3}$ &2.46 $\times 10^{-3}$ \\
{\rm Cs} &133 &55 &7.67 $\times 10^{-3}$ &3.37 $\times 10^{-3}$ \\
{\rm Pt} &195.1 &78 &6.95 $\times 10^{-3}$ &4.09 $\times 10^{-3}$ \\
\end{tabular}
\end{table}

The two main lessons we can draw from Eq. (\ref{approxalphaA}) and the numbers in Table 1 are:
(i) Contrary to what general phenomenological considerations (of the type of Eq. (\ref{compdependence}))
could suggest, there are {\em only two} dominant EP violation effects: one, $Q'_e$, coming from the $\varphi$
sensitivity of the fine-structure constant,  and the other one, $Q'_{\hat m}$, coming from the  $\varphi$
sensitivity of the average light quark mass in nuclear binding;  (ii) in spite of the seemingly small numerical
 coefficient entering the $Q'_e$ term, this term can be comparable to the $Q'_{\hat m}$ one for heavy elements,
such as Platinum or beyond. Actually, one should remember that it is only the {\em variations} of
the $Q$'s over the periodic table which matters. From this point of view, note that the total
variation of $Q'_{\hat m}$ between Li and Pt is $\sim 10^{-2}$, while the corresponding total variation of
$Q'_e$ is  $\sim 4 \times 10^{-3}$. Moreover, while the variation of $Q'_{\hat m}$ is localized
around the light elements, that of $Q'_e$ keeps increasing for heavy elements. [Formally,
 $Q'_{e} \propto Z^2 / A^{4/3} \sim A^{2/3}$, while $Q'_{\hat m} \propto A^{- 1/3}$.]

Summarizing: our theoretical framework suggests that there are two dominant `directions' for the EP-violation signals
associated to a long-range dilaton-like field, namely
\begin{equation}
\left( \frac{\Delta a}{a} \right)_{BC} = (\alpha_B- \alpha_C)\alpha_E =  \left[D_{\hat m} Q'_{\hat m} + D_e Q'_e \right]_{BC}
\end{equation}
where $[Q]_{BC} \equiv Q_B - Q_C$, and where the ` dilaton charges´' are (approximately) given by Eq. (\ref{Qmhat}) and
Eq. (\ref{Qe}). The coefficients $D$ are given by
\begin{equation}
 D_{\hat m} = d^*_g \, (d_{\hat m} - d_g) \, , \qquad D_e = d^*_g \, d_e
\end{equation}
where
\begin{equation}
d^*_g \simeq \alpha^{c.i.} \simeq d_g + 0.093 (d_{\hat m} - d_g)
\end{equation}
If we were assuming that the dilaton coupling $d_e$ is much smaller than $d_{\hat m} - d_g$, we could go further
and conclude (in view of the numerical results indicated in Table I) that the signal $Q'_e$ is sub-dominant
w.r.t.  $Q'_{\hat m}$. In that case we would end up with a uni-dimensional EP signal proportional to $[Q'_{\hat m}]_{BC}$.

\subsection{Experimental bounds}

The fact that two types of EP signals are expected to dominate allow one to derive simultaneous constraints
on the two dominant theoretical parameters $D_{\hat m}, D_e$ by using only two independent sets of EP experiments.
We  can use to that effect the two current EP experiments which have reached the $10^{-13}$ level, namely
the terrestrial E\"{o}tWash experiment, and the celestial Lunar Laser Ranging one

The E\"{o}tWash collaboration has compared the relative
acceleration of Be and Ti in the gravitational field of the Earth \cite{Schlamminger:2007ht}.
The Lunar Laser Ranging (LLR) experiments \cite{lunarlaser} measured the differential acceleration of
the Earth and the Moon towards the Sun. We can use our framework to translate the results from these
two experiments on constraints on the two theoretical parameters $D_{\hat m}, D_e$.

The  E\"{o}tWash result concerns Be (A=9, Z=4) and Ti (A=47.9, Z=22), and reads
\begin{equation}
\left(\frac{\Delta a}{a}\right)_{\rm Be \, Ti} = (\alpha_{Be}-\alpha_{Ti})\alpha_{\rm Earth} = (0.3 \pm 1.8)\times 10^{-13}
\end{equation}
Working at the two-sigma level, i.e. $(0.3 \pm 3.6)\times 10^{-13}$, and neglecting the central value $0.3$,
the rewriting of this equation in terms of the theoretical parameters  $D_{\hat m}, D_e$ yields
 \begin{equation}
 10^{-3} \, \left[ - 7.11 D_{\hat m} - 1.55 D_e \right] = \pm 3.6 \times 10^{-13}
\end{equation}
The Lunar Laser Ranging measurement constrains the relative acceleration of the Earth and the Moon towards the Sun:
\begin{equation}
\left( \frac{\Delta a}{a}\right)_{\rm Earth \, Moon} = (\alpha_{\rm Earth}-\alpha_{\rm Moon})\alpha_{\rm Sun} = ( -1.0 \pm 1.4)\times 10^{-13}
\end{equation}

In addition to the composition dependence of the matter in these objects, it has the remarkable ability to test the
equivalence of the gravitational self energy \cite{Nordtvedt:1968zz}. For dilaton models where the scalar also couples
to matter, it is the matter couplings which will be most important\footnote{Indeed, gravitational self-energy couples
to the combination $\eta_g= 4 (\beta-1) - (\gamma-1)$ of post-Newtonian parameters  \cite{Nordtvedt:1968zz}.
However, this combination is theoretically predicted \cite{Damour:1992we} to be proportional to
$(1-\gamma)/2 \simeq \alpha^{\rm c.i.} \sim d_g^2$ (see above). The fact that the gravitational self energy is a very small fraction
of the total mass then allows one to neglect the corresponding effect.}, and we will not consider here the
gravitational couplings.  The Moon has a very similar composition as the Earth's mantle, which is mostly silicate
(primarily silicon and oxygen). The composition differences between the Earth and the Moon come primarily from the
Earth's core which is dominantly iron.

We approximate the mantle composition as being SiO$_2$, and the Earth' core as being iron. In addition, we follow Ref. \cite{Damour:1995gi} in assigning to the core a relative mass of $32\%$.
Working as above at the 2-sigma level, and rewriting this constraint in terms of our
theoretical parameters\footnote{Strictly speaking one should take into account the fact that the EP signal involves slightly different values for the
`external'  $\alpha_{E}$, namely the Earth versus the Sun. For simplicity, we use here the (justified)
approximation where both are close to the composition-independent part $d^*_g$ of $d_g$.} yields
\begin{equation}
0.32 \times 10^{-3} \, \left[  3.55 D_{\hat m} + 1.0 D_e \right] = \pm 2.8 \times 10^{-13}
\end{equation}
It is interesting to notice the origin of the various numerical coefficients in this equation,
as well as in the corresponding E\"{o}tWash one above. The r.h. sides feature the $ 10^{-13}$
sensitivity level. The l.h.sides have coefficients of order a few times $ 10^{-3}$,
which is typical for the differences of  `dilaton charges' listed in Table I. In addition,
the LLR l.h.s. has an extra factor $0.32$ due to the fact that only $32\%$ of the Earth
differs in composition from the Moon. Finally, we need to solve two linear equations
for the two unknowns  $D_{\hat m}, D_e$ and this introduces an inverse determinant
which will further increase the result for the $D$'s. At the end of the day, if one
denotes $\epsilon_{Eot} = \pm 3.6 \times 10^{-10}$ and
 $\epsilon_{LLR} = \pm 2.8 \times 10^{-10}$ (i.e. the two, random two-sigma
errors multiplied by $10^3$) the solution for $D_{\hat m}, D_e$ reads
\begin{eqnarray}
D_{\hat m} &=& - 0.625 \, \epsilon_{Eot} -3.0 \, \epsilon_{LLR} \, , \nonumber \\
  D_e &=& 2.2\,  \epsilon_{Eot} + 14.0\,  \epsilon_{LLR} \, .
\end{eqnarray}
If $ \epsilon_{Eot}$ and $\epsilon_{LLR}$ were non-zero EP violation signals, this would give us the values of the dilaton parameters in terms of EP data. In the present situation, however, $ \epsilon_{Eot}$ and $\epsilon_{LLR}$ are only (independent) random errors.
This expression then shows that the LLR error is dominating the error level in the
final result. A LLR EP measurement should be about six times below the $ 10^{-13}$ level
to contribute the same error level as a terrestrial EP measurement at the  $ 10^{-13}$ level.
Adding the right-hand-sides of the previous expressions in quadrature, finally leads
to the following (two-sigma) error levels on our theoretical parameters
\begin{equation}
 D_{\hat m} = \pm 0.87 \times 10^{-9}, \, \quad D_e = \pm 4.0 \times 10^{-9}
\end{equation}

\subsection{Specific models}

As we discussed above, one expects that a string-theory dilaton (or moduli) will have low-energy couplings
to matter of the general form $d_g \sim K d_c$, $d_{m_a} - d_g \sim (K_a-K) d_c$, and $d_e \sim d_c$,
where $d_c$ is some common string-scale dimensionless dilaton coupling,
where the enhancement factors $K, K_a$ are expected to be comparable and of order $40$, and where
$d_e$ does not contain any significant enhancement factor. Using the E\"{o}tWash-LLR-derived constraints given in
the preceding section, we then conclude that the string-scale dilaton coupling $d_c$ is constrained to be
$ d_c^2 \lesssim 10^{-9}/ (K |K - K_{\hat{m}}|) \sim 10^{-12}$.

There are two possible attitudes towards this very stringent constraint. One is to conclude that all
the dilaton-like scalar fields of string theory that are massless at tree level must acquire, via loop effects, a large
enough mass to make them invisible in current EP experiments (i.e. $m_{\phi}^{-1} < 0.2$ mm).
A second possibility (suggested in \cite{Damour:1994ya} ) consists in assuming that loop effects (which
depend on the VEV of the dilaton) modify the usual tree-level  dilaton dependence
($\propto \exp (- 2 \varphi)$)
of the various terms entering the string-scale Lagrangian into more complicated functions of $\varphi$,
say  $B_i(\varphi)$, such that these coupling functions reach an extremum
at a special value, say $\varphi_*$ of $\varphi$.
Indeed, under this assumption, Damour and Polyakov \cite{Damour:1994ya} have shown that the cosmological evolution of the universe
drives the VEV of $\varphi$ towards $\varphi_*$, thereby ensuring that the string-scale dilaton
coupling $d_c$, which is proportional to $\partial \ln g_c^2/ \partial \varphi$, is naturally very
small: ``Least Coupling Principle'' (see also Refs. \cite{Damour:1992kf} and \cite{Brax:2010gi}).
More precisely, \cite{Damour:1994ya} showed that, if the extremum is located at a finite field value $\varphi_*$,
cosmological edvolution would reduce an initial dilaton coupling $d_c^{\rm init}$ by a factor
typically\footnote{We assume
here that the curvature parameter $\kappa$ of the dilaton-coupling function $B(\varphi)$ is of order one. See \cite{Damour:1994ya}
for the $\kappa$ dependence of the total cosmological ``attracting factor'' $F_t(\kappa)$.}
of order $ F_t \sim 10^{-9}$. Taking this attracting factor into account then suggests that the present,
late-cosmological-evolution dilaton coupling coefficients are of order
\begin{equation}
d_g \sim d_g - d_{m_a} \sim 40 \, d_e \sim 4 \times 10^{-8} d_c^{\rm init}
\end{equation}

If we insert this result into the EP violation deduced from our results above, say
\begin{equation}
\label{EPviolation}
(\alpha_{Be}-\alpha_{Ti})\alpha_{\rm Earth} \simeq  7 \times 10^{-3} d_g (d_g-d_{\hat{m}})
\end{equation}
we get a rough ``prediction'' for the level of EP violation of the order
\begin{equation}
(\alpha_{Be}-\alpha_{Ti})\alpha_{\rm Earth} \sim  10^{-17}  \, (d_c^{{\rm init}})^2 \, ,
\end{equation}
where $d_c^{\rm init}$ is expected to be of order unity. We note that this result is compatible with
the current experimental tests of the EP, but that several planned improved EP experiments
\cite{microscope,gg,step, cold atoms} will be
able to probe this level of EP violation.

In another version of this dilaton-cosmological-attractor mechanism, the attractor point $\varphi_*$ is
located at infinity in field space (``runaway dilaton'' model \cite{runaway}). This corresponds
to dilaton-dependent couplings of the form
\begin{equation}
B_i(\varphi) = C_i +b_i  e^{-\varphi} +...
\end{equation}
During the cosmological evolution, the dilaton runs towards (the strong-coupling limit) $\varphi = + \infty$, exponentially suppressing its coupling to matter. Studying the effect of this runaway mechanism during slow-roll inflation allowed Ref. \cite{runaway}
to relate the present value of the composition-independent dilaton coupling $\alpha^{\rm c.i.} \simeq d_g$
to the amplitude $ \delta_H \sim 5 \times 10^{-5}$ of density fluctuations generated during inflation.
This leads to
\begin{equation}
\label{DPV}
d_g \simeq \alpha_{\rm c.i.} \sim 3.2\frac{b_F}{cb_\lambda}\delta_H^{4/(n+2)}
\end{equation}
where $n$ denotes the power of the inflaton $\chi$ in the inflationary potential, $V(\chi) \propto \chi^n$.
For instance, in the case of the simplest inflationary potential $V(\chi)= \frac{1}{2} m_\chi^2 \chi^2$,
i.e. $n=2$, the above result leads to
\begin{equation}
d_g^2  \sim 2.5 \times 10^{-8} \left(\frac{b_F}{cb_\lambda}\right)^2
\end{equation}
In view of our present new results, Eq. (\ref{EPviolation}), on the level of EP violation associated to such a composition-independent coupling, this corresponds to
\begin{equation}
(\alpha_{Be}-\alpha_{Ti})\alpha_{\rm Earth} \sim
 2 \times 10^{-10} \left(\frac{b_F}{cb_\lambda}\right)^2
\end{equation}
This is in conflict with the current EP tests, except if one assumes that the combination
of dimensionless parameters $b_F/(cb_\lambda)$ (which was assumed in \cite{runaway} to be of order unity)
happens to be smaller than about $1/30$. In such a model, one would expect to see EP violations
just below the currently tested level.
Alternatively, one might interpret the constraint from current EP tests as suggesting that the
(effective) power of the inflaton in the inflationary potential $V(\chi)$ is less than $n=2$.
For instance, if $n \approx 0$, Eq. (\ref{DPV}) implies $d_g^2 \simeq 6 \times 10^{-17}$,
corresponding to $\Delta a/a \sim 4 \times 10^{-19}$.

Finally, a recent work \cite{Piazza:2010ye} suggests the existence of couplings of a light scalar
which are quite different from the usual string-motivated ones.
In the model of Ref. \cite{Piazza:2010ye} the light scalar couples only to quark mass terms,
through mixing with the Higgs. At tree level, the couplings are
\begin{equation}
d_{mi}= \frac{A}{\kappa m_H^2}
\end{equation}
where $A$ is a very small mixing parameter and $m_H$ is the mass of the Higgs boson. However, integrating out the heavy (t,b,c) quarks (\`a la  \cite{SVZ78,Kaplan})
induces gluonic couplings
\begin{equation}
d_g = \frac{2A}{9\kappa m_H^2}
\end{equation}
The constraint of this model can be then calculated to be
\begin{equation}
\left[ \frac{A}{\kappa m_H^2}\right]^2 < 4.0\times 10^{-10} ~~.
\end{equation}

\section{Experimental sensitivities}

It can be useful to use a well-motivated parameterized theoretical model as a guideline for
comparing the significance, and relative sensitivities, of different experiments.
For instance, the parametrized
post-Newtonian framework \cite{Will:2005va} played a useful role in comparing the
theoretical significance of various {\em composition-independent}
tests of relativistic gravity. Here, we wish to capitalize on the better understanding,
explained above, of the coupling of a generic dilaton-like field to nuclear binding energy
to propose such parametrized frameworks for comparing different {\em composition-dependent}
tests of gravity. Our proposal is intended as an update, or a specification, of previous
similar proposals (see, e.g. \cite{Damour:1994ya,dent}). Actually, our proposal is two-headed.

On the one hand, if we make minimal assumptions, and essentially no approximations, we propose to
parameterize EP violations by means of the matter coupling (\ref{alpha5}), which involves
five  parameters. One of them, $d_g$ (or more accurately $d_g^* = \langle \alpha_A \rangle$)
measures the composition-independent part of the matter coupling, and can, in principle,
be measured by composition-independent gravity tests. The other four parameters,
$d_{\hat{m}} -d_g, d_{\delta m} -d_g, d_{m_e} - d_g, d_e $ are associated with
four different types of EP-violation signals, associated to the four different
`dilaton charges' $Q_{\hat{m}},  Q_{\delta m},  Q_{m_e}, Q_e$, defined in
Eqs. (\ref{Qmhat}), (\ref{Qdeltam}), (\ref{Qme}), and (\ref{Qe}).

On the other hand, we have pointed out that two `directions' of EP violations are
likely to dominate the measured signals. They correspond to the two charges  $Q_{\hat{m}}$
and $ Q_e$, i.e. to the two dilaton parameters $d_{\hat{m}} -d_g$ and $d_e$. For brevity we shall
denote the first one  as
\begin{equation}
d_q \equiv d_{\hat{m}} -d_g
\end{equation}
It measures the dilaton coupling to the ratio $\hat{m}/\Lambda_3$ of the average light-quark mass
to the QCD scale.
We recall that the second one, $d_e$ is associated to the $\varphi$ sensitivity of the fine-structure
constant $\alpha = e^2/(4 \pi)$. In the same approximation that
these charges dominate, we can simplify the expression of the the matter coupling $\alpha_A$
and the corresponding charges, see Eq. (\ref{approxalphaA}), and the  equations following it.
The latter, simplified two-EP-parameter framework\footnote{In all, this
model contains three independent parameters: $d_g$, $d_q$ and $d_e$.
If one could argue that the $\varphi$ sensitivity of $\kappa \hat{m}$ is much
smaller than that of $\kappa \Lambda_3$ one could even consider a
much more special one-parameter guideline
model keeping only $d_g$ and setting to zero the various mass couplings $d_{m_a}$ as well
as $d_e$. In such a model $d_q= -d_g$ would be fixed in terms of $d_g$.
 However, the no-scale supergravity models (and their string realizations)
rather suggest that the $d_{m_a}$'s contain logarithmic amplification
factors which are comparable to the one expected to be present in $d_g$.}
is quite predictive, and could be useful as a guideline for
comparing and/or planning EP experiments.
Let us briefly indicate some consequences of our proposals.

\subsection{Composition independent constraints}

The first useful result in the simplified ``reference dilaton model'' is the expected
ratio between composition-independent effects and composition-dependent ones. As explained above
the former are essentially measured by the Eddington parameter\footnote{Here $d_g$ should
more accurately be replaced by some average $\langle \alpha_A \rangle \equiv d_g^*= d_g + c d_q$,
with a coefficient $c \sim 0.1 $ depending of the average composition of the considered source bodies.}
\begin{equation}
1- \gamma \simeq 2 d_g^2
\end{equation}
while the latter are given, say, by Eq. (\ref{EPviolation}). Note that the numerical
value $7 \times 10^{-3}$ in the latter equation comes from the $Q'_{\hat{m}}$ charge difference
between Be and Ti. We can use instead the maximal difference of $10^{-2}$
corresponding to Be and Pt. This yields the approximate link
\begin{equation}
\frac{\Delta a}{a} \sim 10^{-2} \frac{d_q}{d_g} \frac{1- \gamma}{2}
\end{equation}
Note that, assuming $d_q \sim d_g$, this differs by two orders of magnitude from the link
$\Delta a/a \sim 10^{-4} (1- \gamma)/2 $ estimated in \cite{Damour:1994ya} from
considering as dominant the EM coupling $d_e$ instead of $d_q$. This suggests that current EP tests
correspond to post-Newtonian tests at the level $(1- \gamma)/2 \sim 10^{-11}$, i.e. six
orders of magnitude below the current best post-Newtonian test, namely the Cassini limit
Eq. (\ref{cassini}). [Using the results derived above from combining Eotwash and LLR data,
one actually gets a constraint at the level  $(1- \gamma)/2 \sim 10^{-9}$, where the
loss of a factor $100$ comes from the combination of effects explained above.]

\subsection{Test materials}

Concerning the comparison among the sensitivities of different EP experiments,
we already gave above an example of the use of our
framework (comparison between Eotwash and LLR). Let us also mention another
illustrative example. Note that each EP comparison of a pair of materials,
say $(B,C)$, corresponds, within our simplified framework, to
looking for a signal of the form $ {\bf D} \cdot {\bf Q}_{BC}$,
where $ \bf{D}$ is the two-dimensional vector of dilaton couplings $(D_{\hat m}, D_e)$ ,
and $ \bf{Q}$ a two-dimensional vector of dilaton-charge differences
$(Q'_{\hat m}, Q'_e)_{BC} = (Q'_{\hat m}, Q'_e)_{B}-(Q'_{\hat m}, Q'_e)_{C} $.
For instance, the current best Eotwash comparison concerned Be and Ti,
i.e. (using Table 1) the `charge' vector ${\bf Q}_{Ti \, Be} = (7.11, 1.55) \times 10^{-3}$.
By contrast, the MICROSCOPE experiment plans to use a pair Ti, Pt, which corresponds
to the charge vector  ${\bf Q}_{Pt \, Ti} = (3.33, 2.04) \times 10^{-3}$.
We see that the two choices are nicely complementary in that the former
(using lighter elements) gives more weight to the $\hat{m}$ component of the
EP violation, while the latter (with heavier elements) gives approximately equal weights to the
$\hat{m}$ and $e$ directions.

\subsection{Atomic interferometry}

 Special mention should be given to the sensitivity of EP experiments
based on atomic-interferometer techniques. For instance, Ref. \cite{cold atoms}
mentions the possibility of comparing two isotopes of Rubidium: ($^{85}$Rb,$^{87}$Rb).
In such a case, we wish to warn the reader that one should not blindly use the
formulas that we have derived above, especially the approximate ones for $(Q'_{\hat m}, Q'_e)$.
Indeed, the approximations used to simplify the charges employed the average link (\ref{Z})
between $Z$ and $A$. This approximation is acceptable if one compares
elements that are distant along
the periodic table, but is definitely invalid for isotopes of the same $Z$.
Therefore, one should start from our original, non-approximated expressions
for the charges.

The use of our (`exact') dilaton  charges,
suggests that an EP test comparing ($^{85}$Rb,$^{87}$Rb) would correspond,
in the full four-dimensional space of  $(\hat{m}, \delta m , m_e, e)$ ,
to a charge vector equal to
${\bf Q}_{^{87}Rb \,^{85}Rb } = (-3.3, 3.4, -0.55, -9.2) \times 10^{-5}$.
Note that the components of this vector are  significantly smaller than
those of the charge vectors probed by the other experiments. The dominant
direction is along $e$.
Note also that the $\delta m$ direction now plays
a role as significant as the $\hat{m}$ one, because, besides the binding energies, a crucial
 effect in comparing two isotopes is evidently a change in the number of neutrons.
This also shows that such experiments are complementary to the usual ones,
in that they probe new directions in theory space, though it comes at the cost of
the overall sensitivity.

The atomic interferometer proposal of \cite{mueller} suggests the comparison of $^7$Li and $^{133}$Cs atoms. In contrast to the Rubidium experiment, these elements are well separated in $A,~Z$, and our simplified charges can be used. We find that this comparison is quite sensitive to the dilaton couplings with dilaton charge vector  ${\bf Q}_{Cs \, Li} = (11.2, 3.02) \times 10^{-3}$. While the experimental comparison of dissimilar atoms may be more difficult than the use of related isotopes, the sensitivity to the dilaton couplings is much increased.

Let us also make some further comments relevant for comparing two isotopes which are very close in mass.
Our derivation assumed that the semi-empirical mass
formula was an accurate representation of the binding energies. However, this
mass formula is  an average, which does not always accurately capture local
fluctuations, and notably fluctuations linked to varying $A$ for a fixed $Z$.
In addition, our derivation has neglected the pairing term $- \delta a_p/A^{1/2}$,
as being subdominant. However, this term might become very important if one
were to compare isotopes with mass numbers $A$ differing by an {\em odd} integer.
Indeed, in that case $\delta = \frac{1}{2} \, [(-)^N + (-)^Z]$ {\em changes by one unit}
between the two isotopes, and therefore yields a full contribution $a_p$ to
their mass difference, and thereby also to the dilaton sensitivity.
Actually, we would suggest to try to take advantage of this fact by using
such odd-related isotopes which are likely to have an enhanced sensitivity
to EP violations. [We are aware, however, that this proposal poses both
theoretical challenges (determining the $\varphi$ sensitivity of $a_p$),
and experimental ones (as the two isotopes will have a different Fermi/Bose
statistics, which might undermine the possibility of using accurate,
Bose-Einstein-Condensation-based, techniques).]

\subsection{Other applications}

 Let us also mention that our framework can be straightforwardly applied to comparing
(weak) equivalence principle tests to atomic-clock tests of the dependence of coupling
constants on the gravitational potential. The link between these two types of tests
has been discussed by several authors \cite{Will:2005va,Damour:1997bf,Nordtvedt:2002qe,dent}.
Let us indicate how it is formulated in our notation. The spacetime dependence of the dilaton
field is approximately of the form: $\varphi(x,t)= \varphi_0(t)+\varphi_{\rm loc}(x,t)$,
where $\varphi_0(t)$ is the cosmological value of $\varphi$, and where
\begin{equation}
\varphi_{\rm loc}(x,t)= -\sum_E \alpha_E \frac{G m_E}{r_E} \simeq - \alpha^{\rm c.i.} U(x,t)
\end{equation}
gives the influence of the local matter distribution, in terms of the local
gravitational potential $U$ ($U>0$). In the second expression, we have used the
approximation $\alpha_E \simeq \alpha^{\rm c.i.} = d_g^*$. Combining this result with
our parametrization $k_a(\varphi) = (1+ d_a \varphi) k_a(0)$ of
the $\varphi$ dependence of the various constants $k_a= \hat{m}/\Lambda_3, \delta m/\Lambda_3,
m_e/\Lambda_3, \alpha=e^2/(4 \pi)$, we see that the local gravitational potential influences
the values of the constants $k_a$ measured, say, on the Earth, according to
\begin{equation}
k_a^{\rm loc}= (1-D_a U) k_a(\varphi_0(t))
\end{equation}
where the coefficients $D_a \equiv d_a \alpha^{\rm c.i.}= d_a d_g^*$,
i.e. $D_{\hat m} = d^*_g \, (d_{\hat m} - d_g) \, , \ldots, D_e = d^*_g \, d_e$
are the {\em same} dilaton coefficients that entered our discussion above of the EP tests.
Then, to compute the effect of the seasonally varying $U$ on, say, the
frequencies of atomic clocks, one needs to know the sensitivity of these
frequencies to variations in the $k_a$'s (see \cite{Flambaum:2006ip}).
In particular, the $D_e =\pm 4 \times 10^{-9}$ two-sigma bound derived above
on $D_e$, combined with the yearly variation $\Delta U \simeq 3 \times 10^{-10}$
linked with the Earth's eccentricity, shows that EP tests constrain the yearly variation
of the fine-structure constant on the Earth to be smaller than $1.2 \times 10^{-18}$ (two sigma).
This is about $40$ times smaller than the current best atomic-clock experimental
sensitivity to the variation of $\alpha$ \cite{Rosenband}. Note, however, that clock-comparison
experiments are sensitive to different combinations of the parameters $D_a$ than EP tests
 \cite{Damour:1997bf}. We shall not discuss here the cosmological aspects of the variation
 of constants, which are more model-dependent. For instance, in the context of the dilaton-runaway  model, one can relate the present rate of variation of the `constants' to the (square root
 of the) EP violation level,  see Eq. (3.25) of \cite{runaway}.

Let us finally remark that it would be interesting to use the recent progress (reported
in \cite{chiral,damourdonoghue} and here) about the quark-mass dependence of nuclear binding
to try to derive a well-justified estimate of the quark-mass dependence of the crucial
very low-energy neutron capture resonance $E_r \simeq 0.1$ eV $= 10^{-7}$ MeV of $^{149}$Sm. Indeed,
the analysis of the Oklo data \cite{Damour:1996zw,Gould:2007au,Petrov:2005pu}
shows that this resonance has not changed by more than about $0.1$ eV since the
Oklo natural fission reactor was in activity 2 billion years ago. A naive
use of our results, based on our finding that the bulk binding energy per nucleon,
$a_v$, varies with ${\hat m}$ as $ \Delta a_v \simeq - 42  \Delta \ln {\hat m}/\Lambda_3$ MeV,
suggests that Oklo data constrain the fractional variation of ${\hat m}/\Lambda_3$ over
2 billion years to the level $\Delta \ln {\hat m}/\Lambda_3 \lesssim 10^{-7}/42 \sim 2.4 \times 10^{-9}$. Such a limit would be a very significant constraint on the possible cosmological
evolution of the dilaton. However, it is not clear whether a detailed study of the specific (unstable)
energy level corresponding to $E_r$ will confirm this sensitivity to ${\hat m}/\Lambda_3$.

\section{Conclusions}

We have provided a parametrized framework for the study of the
equivalence principle\footnote{Here we have
limited our considerations to the weak equivalence principle (tests of the universality
of free fall). However, our parametrized Lagrangian can also be used to study the effect
of dilaton couplings on other aspects of the EP: such has clock-comparison experiments.}
in models with light, dilaton-like scalar particles. Our general framework contains
five independent parameters, and should be applicable to the low-energy limit of many models.
The most novel aspect of our work was to provide an estimate of the effects of the dilaton coupling
to nuclear binding energy.  We have found that these couplings  induce, as leading effect, equivalence principle violations
varying with the mass number as $A^{-1/3}$. The level of these EP violations is expected to be at least
comparable to (and, for lighter elements, somewhat larger than) that associated to the
Coulomb energy.

We have also provided a simplified scalar model, containing three parameters: one composition-independent
parameter, and two composition-dependent ones. This model is expected to describe the dominant effects of the
most general 5-parameter framework. We suggest to use it as a guideline for comparing and planning EP experiments.
We used it to combine E\"otvos and Lunar Laser Ranging data so as to constrain its two theoretical composition-dependent
parameters. We found that they are constrained at the $10^{-9}$ level. This plausibly implies (in our model, and
using some naturality assumption) a corresponding limit on composition-independent effects at about the
same level, i.e.  $(1-\gamma)/2 \lesssim 10^{-9}$, which is four orders of magnitude
below the best present composition-independent gravitational tests (Cassini experiment).

In the happy future situation of several non-zero measurements of EP violations, one could check the consistency of
our simplified model, which is quite predictive. If needed the other scalar couplings could readily be included to make sense of subleading
effects modifying the simple predictions of this simplified model.

\section*{Acknowledgements}
JFD thanks the IHES for hospitality both at the start of this project and at its conclusion. He also acknowledges support partially by the NSF grants PHY- 055304 and PHY - 0855119, and in part by the Foundational Questions
Institute. We thank Ulf Mei{\ss}ner for a useful correspondence.

\section{Appendix: The strange quark mass}

 We are not able to provide a definitive calculation of how equivalence principle violations depend on the strange quark couplings.
This is an area where there is no consensus and the motifs of the day change quickly. While we cannot solve this issue, we will here
 argue that the the strange quark dependence could be about or within the uncertainty that we are quoting.

When quarks are heavy, they can be integrated out with the result simply going into a modification of the gluonic coupling, $d_g$.
The $u,~d$ quarks are light, are directly involved in nucleon couplings and are clearly active dynamically in nucleon binding. The strange quark is intermediate in mass. Nucleons do not explicitly contain strange quarks, so their effects are secondary. Certainly they couple to nucleons at some level through loop effects. Initial theoretical calculations suggested that these couplings could be quite large. However, increasingly theoretical and experimental developments are bounding these effects to be relatively small.

Fortunately for equivalence principle violations, the leading manifestations of the strange quark mass would not have an effect in any case. For example, the much debated contribution of the strange quark to the mass of neutrons and protons \cite{strangenucleon} would not lead to the violation of the equivalence principle. This is because the effect is an isospin singlet and contributes equally to the neutron and proton, so that the total effect in an atomic state is proportional to $A$. This leads to a constant contribution to $\alpha_A$ independent of $A$, and no violation of the equivalence principle. Note that the large effects suggested for strange contributions to nucleon masses recently have been bounded by lattice computations to be consistent with zero \cite{JLQCD}. In nuclear binding, the leading $A$ dependent term does not violate the equivalence principle, and it is only the surface term that is relevant. Therefore the key feature to be estimated is the strange quark contribution to the surface binding energy.

In discussing the binding energy it is easy to be led astray. For example early estimates used kaon loops in chiral perturbation theory to conclude that there was a very large effect \cite{olive}. However, it has become clear from dispersive work, such as our own, that the $\bar{K}K$ intermediate state enters above the region of validity of chiral calculations \cite{EFTsfail}. There are analytic studies that show that the reliable low energy portions from such loops are very small \cite{ldr}, and lattice studies have definitively shown that the chiral loop effects are not strongly present at such large masses \cite{Golterman}.

The lightest intermediate states involving strangeness that can couple to nucleons are that of a $K\bar{K}$ intermediate state and also the vector $\varphi$ meson (an $\bar{s}s$ bound state, not to be confused with our notation for the dilaton). In dispersive treatments, both of these start at 1 GeV. The coupling of the $\varphi$ to nucleons is highly uncertain, and depends more on the assumptions made in a given calculation than in a unique piece of evidence in its favor. Moreover it is highly constrained by recent experiments \cite{experiments} that show smaller
than expected hidden strange couplings in nucleons. If we use a estimate which we find to be reasonable and which is within the constraints of present experiments \cite{Meissner:1997qt}, the $\varphi$ effects are too small to be significant.

However, $K\bar{K}$ intermediate states can contribute to the leading scalar interaction and may have a non-trivial effect. We expect from most models of the nuclear potential that most of the scalar strength comes from below 1 GeV. The effect of $M\bar{M}$ intermediate states must decouple as the mass of the meson $M$ gets large. If we estimate generously that $K\bar{K}$ intermediate states contributes $10-15\%$ to the scalar strength, and we take a typical form factor to account for the high mass threshold of the form $(\Lambda^2 + 4m_K^2)^{-1}$ (where $\Lambda$ is is some typical form factor scale) , we would estimate the strange quark mass dependence
\begin{equation}
m_s \frac{\partial \eta_S}{\partial m_s}= m_K^2 \frac{\partial \eta_S}{\partial m_K^2}= (0.10-0.15)\frac{4m_K^2}{\Lambda^2+4m_K^2} \sim 0.07 - 0.10
\end{equation}
using $\Lambda^2 = m_\rho^2$. Comparison with Eq. (\ref{etaS}) indicates that this is comparable to the error bar that we assigned to that calculation.
If the $K\bar{K}$ is positive as expected, a contribution of this size could lead to a 20-30 \% increase in the coefficient of the leading $A^{-1/3}$
 term in our final results. This is clearly a crude estimate, but we don't expect that it is grossly misleading.


\begin{thebibliography}{99}

\bibitem{Schlamminger:2007ht}
  S.~Schlamminger, K.~Y.~Choi, T.~A.~Wagner, J.~H.~Gundlach and E.~G.~Adelberger,
  ``Test of the Equivalence Principle Using a Rotating Torsion Balance,''
  Phys.\ Rev.\ Lett.\  {\bf 100}, 041101 (2008)
  [arXiv:0712.0607 [gr-qc]].

\bibitem{lunarlaser}
  J.~G.~Williams, S.~G.~Turyshev and D.~H.~Boggs,
 ``Progress in Lunar Laser Ranging Tests of Relativistic Gravity,''
  Phys.\ Rev.\ Lett.\  {\bf 93}, 261101 (2004)
  [arXiv:gr-qc/0411113].
J.~G.~Williams, S.~G.~Turyshev and D.~H.~Boggs,
  ``Lunar Laser Ranging Tests of the Equivalence Principle with the Earth and
  Moon,''
  Int.\ J.\ Mod.\ Phys.\  D {\bf 18}, 1129 (2009)
  [arXiv:gr-qc/0507083].



\bibitem{damour}
  T.~Damour,
  ``Questioning the equivalence principle,''
  arXiv:gr-qc/0109063.\\
  T.~Damour,
  ``Testing the equivalence principle: Why and how?,''
  Class.\ Quant.\ Grav.\  {\bf 13}, A33 (1996)
  [arXiv:gr-qc/9606080].
T. Damour, J.P. Blaser;
``Optimizing the choice of materials in equivalence principle experiments"
 in Particle Astrophysics, Atomic Physics and Gravitation, Proceedings of the XIVth Moriond Workshop,
edited by J. Tran Than Van, G. Fontaine and E. Hinds, Editions Frontières, Gif-sur-Yvette, pp 433-440, 1994.
  	  	

\bibitem{Turyshev:2009ir}
  E.~G.~Adelberger {\it et al.},
  ``Opportunities for Probing Fundamental Gravity with Solar System
  Experiments,''
  arXiv:0902.3004 [gr-qc].

\bibitem{microscope}
http://smsc.cnes.fr/MICROSCOPE/index.htm

\bibitem{gg}
http://eotvos.dm.unipi.it/

\bibitem{step}
http://www.sstd.rl.ac.uk/fundphys/step/  \\
http://einstein.stanford.edu/STEP/index.html


\bibitem{cold atoms}
S.~Dimopoulos, P.~W.~Graham, J.~M.~Hogan and M.~A.~Kasevich,
  ``Testing General Relativity with Atom Interferometry,''
  Phys.\ Rev.\ Lett.\  {\bf 98}, 111102 (2007)
  [arXiv:gr-qc/0610047].\\
  S.~Dimopoulos, P.~W.~Graham, J.~M.~Hogan and M.~A.~Kasevich,
  ``General Relativistic Effects in Atom Interferometry,''
  Phys.\ Rev.\  D {\bf 78}, 042003 (2008)
  [arXiv:0802.4098 [hep-ph]].

\bibitem{mueller}
G.~Kim and H.~ Mueller, ``Test of the equivalence principle using Li atom interferometry'', Bull. of the American Physical Society Vol. 55, No 5, E1.00009 (2010)

\bibitem{POEM}
R.~D.~Reasenberg and J.~D.~Phillips,
  ``A weak equivalence principle test on a suborbital rocket,''
  Class.\ Quant.\ Grav.\  {\bf 27} (2010) 095005
  [arXiv:1001.4752 [gr-qc]].



\bibitem{Taylor:1988nw}
  T.~R.~Taylor and G.~Veneziano,
  ``Dilaton Couplings at Large Distances,''
  Phys.\ Lett.\  B {\bf 213}, 450 (1988).



\bibitem{Damour:1994ya}
  T.~Damour and A.~M.~Polyakov,
  ``String theory and gravity,''
  Gen.\ Rel.\ Grav.\  {\bf 26}, 1171 (1994)
  [arXiv:gr-qc/9411069].\\
  T.~Damour and A.~M.~Polyakov,
  ``The String Dilaton And A Least Coupling Principle,''
  Nucl.\ Phys.\  B {\bf 423}, 532 (1994)
  [arXiv:hep-th/9401069].


\bibitem{runaway}
  T.~Damour, F.~Piazza and G.~Veneziano,
  ``Runaway dilaton and equivalence principle violations,''
  Phys.\ Rev.\ Lett.\  {\bf 89}, 081601 (2002)
  [arXiv:gr-qc/0204094].\\
  T.~Damour, F.~Piazza and G.~Veneziano,
  ``Violations of the equivalence principle in a dilaton-runaway scenario,''
  Phys.\ Rev.\  D {\bf 66}, 046007 (2002)
  [arXiv:hep-th/0205111].

\bibitem{Damour:1992we}
  T.~Damour and G.~Esposito-Farese,
  ``Tensor multiscalar theories of gravitation,''
  Class.\ Quant.\ Grav.\  {\bf 9}, 2093 (1992).


\bibitem{Kaplan}
  D.~B.~Kaplan and M.~B.~Wise,
  ``Couplings of a light dilaton and violations of the equivalence
  principle,''
  JHEP {\bf 0008}, 037 (2000)
  [arXiv:hep-ph/0008116].

\bibitem{equivalenceletter}
T.~Damour and J.~F.~Donoghue, ``Phenomenology of the equivalence principle with light scalars'', 
arXiv:1007.2790 [gr-qc].

\bibitem{dent}
  T.~Dent,
  ``Eotvos bounds on couplings of fundamental parameters to gravity,''
  Phys.\ Rev.\ Lett.\  {\bf 101}, 041102 (2008)
  [arXiv:0805.0318 [hep-ph]].


\bibitem{SVZ78}
  M.~A.~Shifman, A.~I.~Vainshtein and V.~I.~Zakharov,
  ``Remarks On Higgs Boson Interactions With Nucleons,''
  Phys.\ Lett.\  B {\bf 78}, 443 (1978).

\bibitem{CollinsDuncanJoglekar77}
  J.~C.~Collins, A.~Duncan and S.~D.~Joglekar,
  ``Trace And Dilatation Anomalies In Gauge Theories,''
  Phys.\ Rev.\  D {\bf 16}, 438 (1977).

\bibitem{RPP}
C. Amsler et al. (Particle Data Group), Physics Letters B667, 1 (2008) and
2009 partial update for the 2010 edition

\bibitem{chiral}
  J.~F.~Donoghue,
  ``The Nuclear Central Force in the Chiral Limit,''
  Phys.\ Rev.\  C {\bf 74}, 024002 (2006)
  [arXiv:nucl-th/0603016].


\bibitem{damourdonoghue}
  T.~Damour and J.~F.~Donoghue,
  ``Constraints on the variability of quark masses from nuclear binding,''
  Phys.\ Rev.\  D {\bf 78}, 014014 (2008)
  [arXiv:0712.2968 [hep-ph]].


\bibitem{massformula}
  P.~E.~Hodgson, E.~Gadioli and E.~Gadioli Erba,
  ``Introductory nuclear physics,''
{\it  Oxford, UK: Clarendon (1997) 723 p}


\bibitem{furnstahl}
R.~J.~Furnstahl and B.~D.~Serot, ``Parameter Counting in
Relativistic Mean-Field Models,'' Nucl.\ Phys.\ A {\bf 671}, 447
(2000) [arXiv:nucl-th/9911019].

\bibitem{contact}
B.~A Nilolaus, T. Hoch and D.~G. Madland, ``Nuclear ground state
properties in a relativistic point coupling model'', Phys. Rev.
{\bf C46}, 1757 (1992)  \\
 B.~D.~Serot and
J.~D.~Walecka, ``Effective field theory in nuclear
many-body physics,'' arXiv:nucl-th/0010031.\\
 J.~J.~Rusnak and R.~J.~Furnstahl,
``Relativistic point-coupling models as effective theories of
nuclei,'' Nucl.\ Phys.\ A {\bf 627}, 495 (1997)
[arXiv:nucl-th/9708040].\\
B. Machleidt and D. R. Entem, ``Towards a consistent approach to
nuclear structure: EFT of two- and many-body forces''
arXive:nucl-th/0503025

\bibitem{sigma}
J.~F.~Donoghue,
  ``Sigma exchange in the nuclear force and effective field theory,''
  Phys.\ Lett.\  B {\bf 643}, 165 (2006)
  [arXiv:nucl-th/0602074].



\bibitem{twonucleon}
  S.~R.~Beane and M.~J.~Savage,
  ``Variation of fundamental couplings and nuclear forces,''
  Nucl.\ Phys.\  A {\bf 713}, 148 (2003)
  [arXiv:hep-ph/0206113].\\
  S.~R.~Beane and M.~J.~Savage,
  ``The quark mass dependence of two-nucleon systems,''
  Nucl.\ Phys.\  A {\bf 717}, 91 (2003)
  [arXiv:nucl-th/0208021].\\
  E.~Epelbaum, U.~G.~Meissner and W.~Gloeckle,
  ``Further comments on nuclear forces in the chiral limit,''
  arXiv:nucl-th/0208040.
 V.~V.~Flambaum and R.~B.~Wiringa,
  ``Dependence of nuclear binding on hadronic mass variation,''
  Phys.\ Rev.\  C {\bf 76}, 054002 (2007)
  [arXiv:0709.0077 [nucl-th]].

\bibitem{eft}
E.~Epelbaum, W.~Glockle, A.~Kruger and U.~G.~Meissner, ``Effective
theory for the two-nucleon system,'' Nucl.\ Phys.\ A {\bf 645}, 413
(1999) [arXiv:nucl-th/9809084].\\
M.~J.~Savage, ``Effective field theory for nuclear physics,''
arXiv:nucl-th/0301058.\\
P.~F.~Bedaque and U.~van Kolck, ``Effective field theory for
few-nucleon systems,'' Ann.\ Rev.\ Nucl.\ Part.\ Sci.\  {\bf 52},
339 (2002) [arXiv:nucl-th/0203055].

\bibitem{dispersive}
E.~Epelbaum, W.~Gl\"{o}ckle and U.~G.~Mei{\ss}ner, ``Improving the
convergence of the chiral expansion for nuclear forces.  I:
Peripheral phases,'' arXiv:nucl-th/0304037. \\
E.~Epelbaum, W.~Gloeckle and U.~G.~Mei{\ss}ner, ``Improving the
convergence of the chiral expansion for nuclear forces.  II: Low
phases and the deuteron,''
  Eur.\ Phys.\ J.\ A {\bf 19}, 401 (2004)
  [arXiv:nucl-th/0308010].\\
N.~Kaiser, ``Chiral 2$\pi$ exchange N N potentials: Two-loop
contributions,'' Phys.\ Rev.\ C {\bf 64}, 057001 (2001)
[arXiv:nucl-th/0107064].\\
 D.~R.~Entem and R.~Machleidt,
  ``Chiral 2$\pi$ exchange at order four and peripheral N N scattering,''
  Phys.\ Rev.\ C {\bf 66}, 014002 (2002)
  [arXiv:nucl-th/0202039].
N.~Kaiser, ``Chiral 2$\pi$-exchange N N potentials: Relativistic
$1/M^2$-corrections,'' Phys.\ Rev.\ C {\bf 65}, 017001 (2002)
[arXiv:nucl-th/0109071].\\
 N.~Kaiser, ``Chiral 3$\pi$ exchange N N
potentials: Results for representation-invariant classes of
diagrams,'' Phys.\ Rev.\ C {\bf 61}, 014003 (2000)
[arXiv:nucl-th/9910044].\\
N.~Kaiser, ``Chiral 3$\pi$ exchange N N potentials: Results for
diagrams proportional to $g_A^4$ and $g_A^6$,'' Phys.\ Rev.\ C {\bf
62}, 024001 (2000) [arXiv:nucl-th/9912054]. \\
N.~Kaiser, ``Chiral 3$\pi$-exchange N N potentials: Results for
dominant next-to-leading order contributions,'' Phys.\ Rev.\ C {\bf
63}, 044010 (2001) [arXiv:nucl-th/0101052].\\


\bibitem{serot}
  B.~D.~Serot and J.~D.~Walecka,
  ``Recent progress in quantum hadrodynamics,''
  Int.\ J.\ Mod.\ Phys.\  E {\bf 6}, 515 (1997)
  [arXiv:nucl-th/9701058].





\bibitem{quark masses}
  J.~Gasser and H.~Leutwyler,
  ``Quark Masses,''
  Phys.\ Rept.\  {\bf 87}, 77 (1982).

\bibitem{Bertotti:2003rm}
  B.~Bertotti, L.~Iess and P.~Tortora,
  ``A test of general relativity using radio links with the Cassini
  spacecraft,''
  Nature {\bf 425} (2003) 374.

\bibitem{Gasser:1990ce}
  J.~Gasser, H.~Leutwyler and M.~E.~Sainio,
  ``Sigma term update,''
  Phys.\ Lett.\  B {\bf 253}, 252 (1991).





\bibitem{Nordtvedt:1968zz}
  K.~Nordtvedt,
  ``Testing relativity with laser ranging to the moon,''
  Phys.\ Rev.\  {\bf 170}, 1186 (1968).









\bibitem{earth}
D.L.~Anderson, {\em Theory of the Earth} (Blackwell Scientific Publications, Oxford, 1989), p.18, cited in S.~Baessler, B.~R.~Heckel, E.~G.~Adelberger, J.~H.~Gundlach, U.~Schmidt and H.~E.~Swanson,
  ``Improved Test of the Equivalence Principle for Gravitational Self-Energy,''
  Phys.\ Rev.\ Lett.\  {\bf 83}, 3585 (1999).


\bibitem{Damour:1995gi}
  T.~Damour and D.~Vokrouhlicky,
  ``The Equivalence Principle And The Moon,''
  Phys.\ Rev.\  D {\bf 53}, 4177 (1996)
  [arXiv:gr-qc/9507016].

\bibitem{Ellis:1984bm}
  J.~R.~Ellis, C.~Kounnas and D.~V.~Nanopoulos,
  ``No Scale Supersymmetric Guts,''
  Nucl.\ Phys.\  B {\bf 247}, 373 (1984).

\bibitem{Damour:1992kf}
  T.~Damour and K.~Nordtvedt,
  ``General relativity as a cosmological attractor of tensor scalar theories,''
  Phys.\ Rev.\ Lett.\  {\bf 70}, 2217 (1993);
  T.~Damour and K.~Nordtvedt,
 and  ``Tensor - scalar cosmological models and their relaxation toward general
  relativity,''
  Phys.\ Rev.\  D {\bf 48}, 3436 (1993).

\bibitem{Brax:2010gi}
  P.~Brax, C.~van de Bruck, A.~C.~Davis and D.~J.~Shaw,
  ``The Dilaton and Modified Gravity,''
  arXiv:1005.3735 [astro-ph.CO].


\bibitem{Piazza:2010ye}
  F.~Piazza and M.~Pospelov,
  ``Sub-eV scalar dark matter through the super-renormalizable Higgs portal,''
  arXiv:1003.2313 [hep-ph].

\bibitem{Will:2005va}
  C.~M.~Will,
  ``The confrontation between general relativity and experiment,''
  Living Rev.\ Rel.\  {\bf 9}, 3 (2005)
  [arXiv:gr-qc/0510072].





\bibitem{Damour:1997bf}
  T.~Damour,
  ``Gravity, equivalence principle and clocks,''
  arXiv:gr-qc/9711060;
  ``Equivalence principle and clocks,''
  arXiv:gr-qc/9904032.

\bibitem{Nordtvedt:2002qe}
  K.~Nordtvedt,
  ``Space-time variation of physical constants and the equivalence principle,''
  Int.\ J.\ Mod.\ Phys.\  A {\bf 17}, 2711 (2002).

\bibitem{Flambaum:2006ip}
  V.~V.~Flambaum and A.~F.~Tedesco,
  ``Dependence of nuclear magnetic moments on quark masses and limits on
  temporal variation of fundamental constants from atomic clock  experiments,''
  Phys.\ Rev.\  C {\bf 73}, 055501 (2006)
  [arXiv:nucl-th/0601050].

\bibitem{Rosenband}
 T. Rosenband et al.
 Frequency Ratio of Al+ and Hg+ Single-Ion Optical Clocks; Metrology at the 17th Decimal Place
 Science {\bf 319}, 1808 (2008). [DOI: 10.1126/science.1154622]

\bibitem{Damour:1996zw}
  T.~Damour and F.~Dyson,
  ``The Oklo bound on the time variation of the fine-structure constant
  revisited,''
  Nucl.\ Phys.\  B {\bf 480}, 37 (1996)
  [arXiv:hep-ph/9606486].

\bibitem{Gould:2007au}
  C.~R.~Gould, E.~I.~Sharapov and S.~K.~Lamoreaux,
  ``Time-variability of alpha from realistic models of Oklo reactors,''
  Phys.\ Rev.\  C {\bf 74}, 024607 (2006)
  [arXiv:nucl-ex/0701019].

\bibitem{Petrov:2005pu}
  Yu.~V.~Petrov, A.~I.~Nazarov, M.~S.~Onegin, V.~Y.~Petrov and E.~G.~Sakhnovsky,
  ``Natural nuclear reactor Oklo and variation of fundamental constants. I:
  Computation of neutronic of fresh core,''
  Phys.\ Rev.\  C {\bf 74}, 064610 (2006)
  [arXiv:hep-ph/0506186].






\bibitem{strangenucleon}
  C.~A.~Dominguez and P.~Langacker,
  ``Present Status Of The Pion - Nucleon Sigma Term,''
  Phys.\ Rev.\  D {\bf 24}, 1905 (1981).\\
  J.~F.~Donoghue and C.~R.~Nappi,
  ``The Quark Content Of The Proton,''
  Phys.\ Lett.\  B {\bf 168}, 105 (1986).\\
 R.~L.~Jaffe and C.~L.~Korpa,
  ``The Pattern of Chiral Symmetry Breaking and the Strange Quark Content of
  the Proton,''
  Comments Nucl.\ Part.\ Phys.\  {\bf 17}, 163 (1987).



\bibitem{JLQCD}
  K.~Takeda, S.~Aoki, S.~Hashimoto, T.~Kaneko, T.~Onogi and N.~Yamada  [JLQCD
                  collaboration],
  ``Calculation of nucleon strange quark content with dynamical overlap
  quarks,''
  arXiv:0910.5036 [hep-lat].\\
 H.~Ohki {\it et al.},
  ``Nucleon sigma term and strange quark content from lattice QCD with exact
  chiral symmetry,''
  Phys.\ Rev.\  D {\bf 78}, 054502 (2008)
  [arXiv:0806.4744 [hep-lat]].



\bibitem{olive}
  K.~A.~Olive, M.~Pospelov, Y.~Z.~Qian, A.~Coc, M.~Casse and E.~Vangioni-Flam,
  ``Constraints on the variations of the fundamental couplings,''
  Phys.\ Rev.\  D {\bf 66}, 045022 (2002)
  [arXiv:hep-ph/0205269].

\bibitem{EFTsfail}
  J.~F.~Donoghue,
  ``When Effective Field Theories Fail,''
  arXiv:0909.0021 [hep-ph].

\bibitem{ldr}
  J.~F.~Donoghue, B.~R.~Holstein and B.~Borasoy,
  ``SU(3) baryon chiral perturbation theory and long distance
  regularization,''
  Phys.\ Rev.\  D {\bf 59}, 036002 (1999)
  [arXiv:hep-ph/9804281].\\
  J.~F.~Donoghue and B.~R.~Holstein,
  ``Improving the convergence of SU(3) baryon chiral perturbation theory,''
  arXiv:hep-ph/9803312.



\bibitem{Golterman}
  M.~Golterman,
  ``Applications of chiral perturbation theory to lattice QCD,''
  arXiv:0912.4042 [hep-lat].



\bibitem{experiments}
   D.~Androic {\it et al.}  [G0 Collaboration],
  ``Strange Quark Contributions to Parity-Violating Asymmetries in the Backward
  Angle G0 Electron Scattering Experiment,''
  Phys.\ Rev.\ Lett.\  {\bf 104}, 012001 (2010)
  [arXiv:0909.5107 [nucl-ex]].\\
S.~Baunack {\it et al.},
  ``Measurement of Strange Quark Contributions to the Vector Form Factors of
  the Proton at Q**2=0.22 (GeV/c)**2,''
  Phys.\ Rev.\ Lett.\  {\bf 102}, 151803 (2009)
  [arXiv:0903.2733 [nucl-ex]].\\
A.~Acha {\it et al.}  [HAPPEX collaboration],
  ``Precision Measurements of the Nucleon Strange Form Factors at Q**2 ~
  0.1-GeV**2,''
  Phys.\ Rev.\ Lett.\  {\bf 98}, 032301 (2007)
  [arXiv:nucl-ex/0609002].\\
  D.~T.~Spayde {\it et al.}  [SAMPLE Collaboration],
  ``The strange quark contribution to the proton's magnetic moment,''
  Phys.\ Lett.\  B {\bf 583}, 79 (2004)
  [arXiv:nucl-ex/0312016].\\
 K.~S.~Kumar and P.~A.~Souder,
  ``Strange quarks and parity violation,''
  Prog.\ Part.\ Nucl.\ Phys.\  {\bf 45}, S333 (2000).


\bibitem{Meissner:1997qt}
  U.~G.~Mei{\ss}ner, V.~Mull, J.~Speth and J.~W.~van Orden,
  ``Strange vector currents and the OZI-rule,''
  Phys.\ Lett.\  B {\bf 408}, 381 (1997)
  [arXiv:hep-ph/9701296].




\end{thebibliography}
\end{document}